\documentclass{article}[12pt]
\usepackage{amsmath}
\usepackage{latexsym}
\usepackage{float}
\usepackage{amssymb}
\usepackage{graphicx}
\usepackage[draft]{pdfpages}
\usepackage{tabularx}
\usepackage{microtype}
\usepackage{etoolbox}
\robustify{\textbf}

\begin{document}

\title{Heterogeneous Nucleation of a Droplet Pinned at a Chemically Inhomogeneous Substrate:\\
A Simulation Study of the Two-dimensional Ising Case}
\author{
               Marta L. Trobo \\[\baselineskip]
               \textit{Instituto de F\'isica de L\'iquidos y Sistemas Biol\'ogicos (IFLYSIB). CCT-CONICET La Plata,}\\
                \textit{UNLP. Calle 59 Nro. 789, (1900) La Plata, Argentina and}\\
                 \textit {Departamento de Ciencias B\'asicas, Facultad de Ingenier\'ia,}
                   \textit{Universidad Nacional de La Plata, Argentina.}\\[\baselineskip]
                                      Ezequiel V. Albano \\[\baselineskip]
                               \textit{Instituto de F\'isica de L\'iquidos y Sistemas Biol\'ogicos (IFLYSIB). CCT-CONICET La Plata,}\\
                \textit{UNLP. Calle 59 Nro. 789, (1900) La Plata, Argentina and}\\
                 \textit {Departamento de F\'isica, Facultad de Ciencias  Exactas,}
                                  \textit{Universidad Nacional de La Plata, Argentina.}\\[\baselineskip]
                       Kurt Binder\\[\baselineskip]
                                 \textit{Institut f\"{u}r Physik, Johannes Gutenberg-Universit\"{a}t Mainz}\\
                 \textit {Staudinger Weg 7, D-55099 Mainz, Germany,}}
\date{\today}



\begin{abstract}
  Heterogeneous nucleation is studied by Monte Carlo simulations and
  phenomenological theory, using the two-dimensional lattice gas model
  with suitable boundary fields. A chemical inhomogeneity of length
  $b$ at one boundary favors the liquid phase, while elsewhere the
  vapor is favored.  Switching on the bulk field $H_b$ favoring the
  liquid, nucleation and growth of the liquid phase starting from
  the region of the chemical inhomogeneity is analyzed. Three regimes
  occur: for small fields, $H_b < H_b^{crit}$, the critical droplet
  radius is so large that a critical droplet having the
  contact angle $\theta_c$ required  by  Young's equation in the
  region of the chemical inhomogeneity
  does not  yet  “fit” there, since the baseline length of the
  circle-cut sphere droplet would exceed $b$.
  For $ H_b^{crit} < H_b <H_b^{\ast}$, such droplets fit inside
  the inhomogeneity, and are indeed
  found in simulations with large enough observation times, but
  these droplets remain pinned to the chemical inhomogeneity when
  their baseline has grown to the length $b$. Assuming that these
  pinned droplets have a circle cut shape and effective contact
  angles $\theta_{eff}$ in the regime $ \theta_c < \theta_{eff} < \pi/2$,
  the density excess due to these droplets can be predicted, and is
  found to be in reasonable agreement with the simulation results.
  On general grounds one can predict that the effective contact
  angle $\theta_{eff}$ as well as the excess density of the droplets,
  scaled by $b^2$, are functions of the product $ b H_b$, but do
  not depend on both variables separately. Since the free energy
  barrier for the “depinning” of the droplet (i.e., growth of
  $\theta_{eff}$ to $\pi - \theta_c$ ) vanishes when $\theta_{eff}$
  approaches  $\pi/2$, in practice  only angles $\theta_{eff}$ up
  to about $\theta_{eff}^{max} \simeq 70^{\circ}$  were observed.
  For larger fields ($H_b > H_b^{*}$) the droplets nucleated at the chemical inhomogeneity
  grow to the full system size. While the relaxation time for the
  growth scales as $\tau_G  \propto H_b^{-1}$, the nucleation
  time $\tau_N$ scales as $\ln \tau_N \propto H_b^{-1}$. However,
  the prefactor in the latter relation, as evaluated for our
  simulations results, is not in accord with an extension of the
  Volmer-Turnbull theory to two-dimensions, when the theoretical
  contact angle $\theta_c$ is used.
\end{abstract}

\maketitle
\newpage
\section{INTRODUCTION}

When thermodynamic variables such as temperature $T$, pressure $p$, or
external fields (e.g. a magnetic field $H$) are varied, discontinuous 
changes in the state of matter can occur. Examples of such so called 
first-order phase transitions \cite{1h,2h} range from the condensation 
of water, melting of ice, crystal formation in solidifying melts, etc., 
to the magnetization reversal of ferromagnetic devices. These phenomena 
are of great importance not only for condensed matter physics, but also 
for the atmospheric sciences, geosciences and material science, as well 
as for numerous technical applications. However, a common feature of 
all these phase changes is that they are triggered by nucleation events, 
i.e. on the background of the old (metastable) phase a nanoscopically small 
nucleus of the new phase needs to be formed, and such nucleation phenomena 
are rare events since a high free energy barrier needs to be 
crossed \cite{2h,3h,4h,5h,6h}. Actually, for most conditions of practical 
interest the spontaneous formation of nuclei by statistical fluctuations, 
i.e the so called "homogeneous nucleation", involves too high barriers and 
cannot happen. In contrast, "heterogeneous nucleation" at defects, 
e.g. condensation of water droplets on dust particles in the atmosphere, 
or surface-induced crystallization starting at the walls of a container, etc., 
occurs much more frequently.
Also, processes such as the formation of dew droplets on 
car windows or plants are familiar from everyday life \cite{6hb}.
However, the nanoscopic size of the nucleus, which 
typically contains only a few hundred of atoms or molecules, is a stumbling
block already for the theoretical description of homogeneous 
nucleation \cite{2h,3h,4h,5h,6h}. The large variety of defects that can 
cause heterogeneous nucleation makes a comprehensive description even more 
difficult, see e.g. \cite{7h}. Thus theoretical work on heterogeneous
nucleation is rather 
scarce \cite{7hb,8h,9h,10h,11h,12h,13h,14h,15h,16h,17h,18h,19h,20h,21h,22h,23h,24h,25h,26h,27h,28h,29h,30h}, 
while theoretical work on homogeneous nucleation is abundant, see 
e.g.\cite{31h} for an overview of work done for Ising/lattice gas models.

In this paper, we reconsider the problem of heterogeneous nucleation on
flat substrates by focusing on a chemically inhomogeneous surface, where 
nucleation preferentially occurs in a region of finite (nanoscopic) linear dimension
$b$, c.f. Figure  \ref{fig1}. There are several motivations for such a choice: 
i) chemically structured surfaces are useful for many applications in
nanotechnology such as the fabrication of nanodevices, for the processing 
of nanoscopic amounts of matter ("lab in a chip"), etc. \cite{32h,33h}. 
ii) Since $b$ is comparable to the size of the nucleus formed in a single 
heterogeneous nucleation event, it is straightforward to study the
characteristics of such isolated nucleation events by computer simulation in 
great detail. In contrast, when one studies nucleation in a system with a 
homogeneous macroscopic surface, surface-attached nuclei can occur anywhere on 
this substrate, and one easily reaches conditions where several nuclei form 
and compete during their growth. This case is familiar from studies of
homogeneous nucleation (e.g. \cite{3h,31h,34h,36h,37h}) and observing 
the lifetime of metastable states in this limit of multi-nuclei formation 
and growth allows only rather indirect conclusions on the nucleation rates 
and nucleation barriers. 
iii) Due to the fact that dust or soot particles at which nucleation happens 
in the atmosphere are often of $\mu m$ size only and have irregular shapes and 
need not be chemically homogeneous, it is of practical interest to study cases 
where conditions favorable for heterogeneous nucleation are limited 
to regions of nanoscopic extent. Of course, also other types of localized
defects are suitable to study isolated heterogeneous nucleation events, 
e.g. in square lattices with free boundary conditions nucleation starts 
at the corners of the square \cite{37hb}.
 
Of course, for this problem there occurs a challenging interplay between 
surface effects due to the substrate and interfacial effects of the material 
forming the nucleus, also statistical fluctuations and finite-size effects 
play a role. Therefore, the development of an analytical theory for the 
treatment of such problems is very difficult \cite{38h,39h,40h}. 
In the present work, we hence restrict our attention to an approach by 
Monte Carlo simulation \cite{41h} of a simple model, namely the Ising/lattice 
gas model. As has been discussed elsewhere\cite{31h}, even this simple model 
presents severe difficulties due to the incomplete knowledge of the 
anisotropy of the interfacial tension between bulk coexisting phases. 
Also understanding of wetting phenomena, e.g. 
contact angles of macroscopic sessile droplets at walls \cite{42h}, 
is a problem in the three-dimensional case (see e.g. \cite{43h,44h,45h}). 
Thus, we focus here on the lattice gas/Ising model in $d = 2$ dimensions, 
for which both bulk and interfacial phenomena including wetting 
behavior \cite{46h,47h,48h,49h,50h,51h,52h} are well understood. 
Recently, we have already studied droplets pinned at chemically heterogeneous 
substrates under conditions of bulk phase coexistence \cite{53h}. 
This knowledge also is useful for the present work where we consider such
droplets under out-of-equilibrium conditions, and the dynamics of the
resulting growth process during nucleation events.

The outline of our paper is as follows:
in Section II, we precisely characterize the model and the simulation method, 
in Section III, we present our results for wall-attached precursors droplets, 
which are in metastable equilibrium during the "observation time" of the
simulation. Furthermore, we discuss a scaling description in terms of the 
variables $b$ (spatial extent of the inhomogeneity) and bulk magnetic 
field $H_b$ (which characterizes the "distance" of the metastable state from 
phase coexistence that occurs in the bulk at $H_b = 0$, of course). 
In Section IV, we present a phenomenological theory of pinned metastable droplets,
which have a baseline $b$ and a non-equilibrium contact angle controlled by the
bulk magnetic field. We discuss the stability limit of these droplets, where they depin
from the chemical inhomogeneity and grow beyond it, causing a fast phase transformation.
In Section V, we analyze the dynamics of nucleation events, characterizing 
both the growth process of a single nucleus from nanoscopic to macroscopic
sizes, and the distribution of nucleation times. Also, we compare these
results to previous findings for single-droplet nucleation in the 
bulk \cite{36h,54h,35h}. 
Finally, Section VI summarizes our conclusions. 
The extension of the classical theory of heterogeneous nucleation 
of Volmer and Turnbull \cite{7hb,8h,9h,10h} to the two-dimensional case is given in Appendix A.
In Appendix B the depinning of droplets from a chemical heterogeneity in $d = 3$ dimensions is
briefly discussed.

\section{Model and simulation details}

\begin{figure}[ht]
\includegraphics[width=10cm,height=6cm,clip]{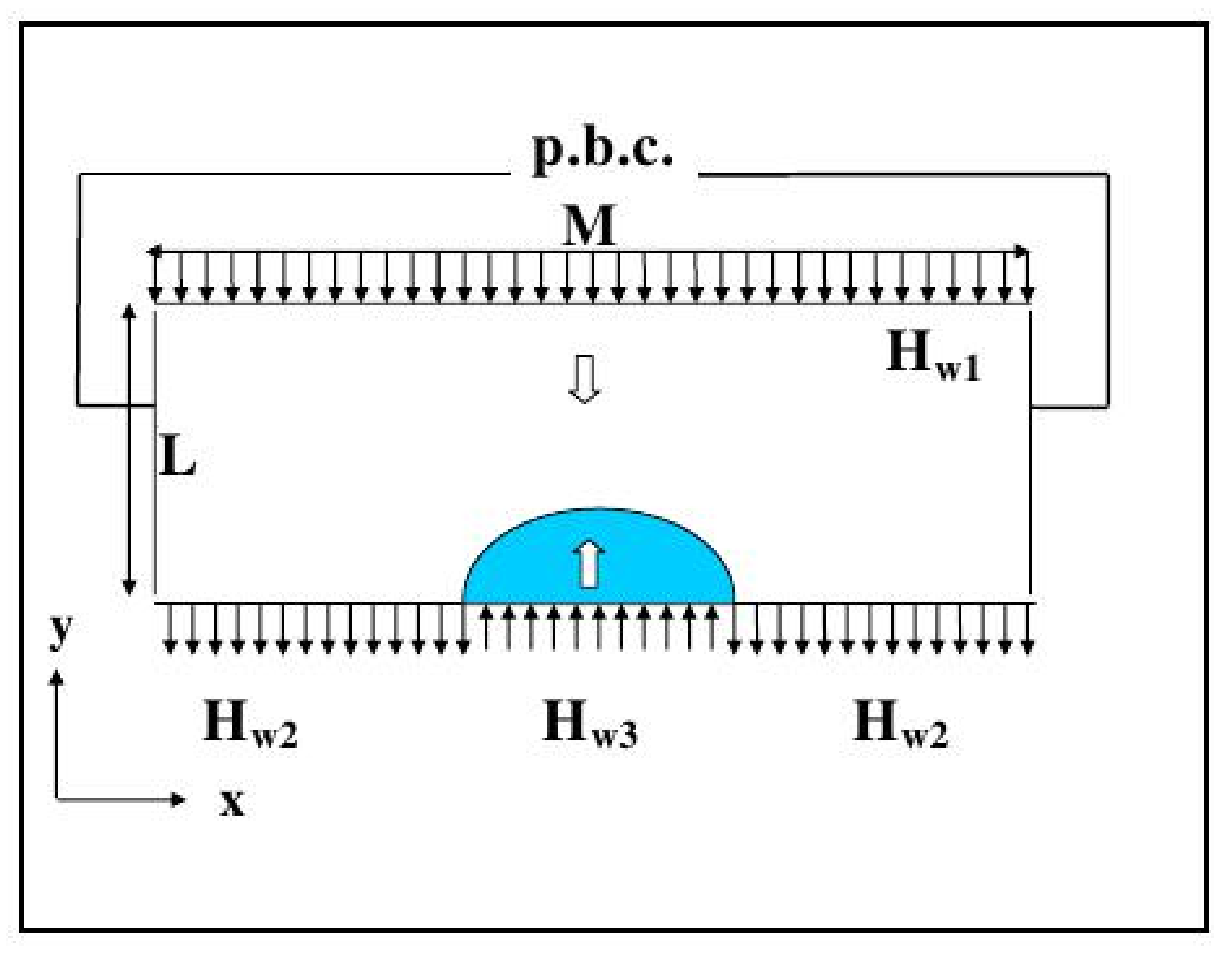}
\caption{\label{fig1} Schematic description of the system geometry. We choose a rectangular 
$M \times L$ lattice with periodic boundary conditions in the $x$-direction, and surface fields acting on the 
first and last row of spins in the $y$-direction. The sign of the surface fields
is indicated by the arrows.  
The surface field on the top row $(H_{w1}$) and outside 
the inhomogeneity at the bottom row $(H_{w2})$ are both chosen negative, so that a negative magnetization 
in the bulk of the system is stable, (thus this area is left in white and the wide arrow indicates the 
negative magnetization). The strength $H_{w1}$ of the surface field on top is chosen 
smaller $(|H_{w1}|=|H_{w2}|/4)$, but for the large linear dimensions chosen the precise 
choice of $H_{w1}$ does not matter with respect to the properties of the droplet.
Also, for $b$ sites at the bottom row a positive field $H_{w3}=|H_{w2}|$ 
is chosen, and a positive bulk field is applied throughout the sample that favors
the heterogeneous nucleation of the droplets within the wall inhomogeneity
(the droplet area is colored and surround
a wide arrow pointing up showing the prevalent positive magnetization in the pinned droplet).
}
\end{figure}

The chosen model is similar to our previous work \cite{53h} where a two-dimensional
Ising/lattice gas model on the square lattice in $L \times M$ geometry was
studied at phase coexistence (bulk field $H_b =0$). We apply periodic boundary
conditions in $x-$direction only, while free boundaries are used in
$y-$direction, and the Ising spins in the first ($i = 1$) and last
($i = L$) rows experience boundary fields, c.f. Figure  \ref{fig1}.
At the upper boundary, we choose a 
homogeneous boundary field $H_{w1} = -0.225$ throughout, which hence favors
the minus spins, $S(i,j) = -1$; in lattice gas language, the local density
variable $\rho(i,j) = (1 + S(i,j))/2$ at a lattice site with coordinates 
$i$ ($i=1,...,L$) in $y-$direction and $j$ ($j=1,...,M$) in $x-$direction then is
$\rho(i,j) = 0$. At the lower boundary, $i=1$, we choose the boundary field 
inhomogeneous: from the site $j = (M-b)/2 + 1$ to the site 
$j = (M+b)/2$ we choose a positive boundary field $H_{w3} = +0.90$, to favor
along a line of length $b$ ($b$ is a large odd integer) the liquid phase.
Outside this region, the boundary field in the first row is $H_{w2} = -0.90$
throughout, to ensure for the chosen total linear dimensions that in the
absence of a bulk field $H_b$ the vapor phase (or phase with negative
magnetization, respectively) is the stable phase, for all choices of $b$ 
that were considered.

The temperature $T$ is measured in units of the critical temperature $T_{cb}$ of the bulk; 
i.e. \cite{46h} $k_BT_{cb}/J=2/\ln (1+ \sqrt{2}) \approx 2.27$, where $k_B$ is Boltzmann's constant, 
and $J$ the exchange constant, respectively. Boundary and bulk fields $H_{w1}$, $H_{w2}$, $H_{w3}$,
$H_b$ are measured in units of $J$. Thus, the used Hamiltonian is

\begin{eqnarray} \label{eqh1}
&& \mathcal{H} =-J \sum\limits_{i=1}^L \sum\limits_{j=1}^M S \Big(i,j\Big) \Big[S\Big(i+1,j\Big) + S\Big(i-1,j\Big) + S\Big(i,j +1 \Big) + S\Big(i, j-1 \Big) \Big]/2 \nonumber\\
&&-H_{w1} \sum\limits_{j=1}^M S \Big(L,j \Big) - \sum\limits^M_{j=1} H_w \Big(j \Big) S\Big(1,j \Big) -
H_b \sum\limits_{i=1}^L \sum\limits_{j=1}^M S \Big(i,j\Big) \quad ; \quad S \Big(i,j \Big) = \pm1 \quad,
\end{eqnarray}

\noindent where $H_w(j) = H_{w2}$ for $1 \leq j \leq (M-b)/2$ and $(M+b)/2 + 1 \leq j \leq M$, while $H_w (j)=H_{w3}$ 
for $(M-b)/2 + 1 \leq j \leq (M +b)/2$. Also, $S(i,j)=0$ is taken for missing neighbors.

The chosen value of $H_{w3}=0.90$ leads to a wetting critical temperature 
$t_w=T_w/T_{cb} \simeq 0.4866$ \cite{48h}. Furthermore heterogeneous nucleation 
is studied for $T < T_w$, so that the order within the bulk domains 
is almost perfect, and the correlation length in the bulk is of the order of the lattice
spacing. For all temperatures $T < T_w$ we have a nonzero contact angle $\theta_c < 90^{\circ}$
in the region where $H_{w3}$ acts, while due to the antisymmetric choice $H_{w2} = -H_{w3}$
the contact angle is $\pi -\theta_c$ in the region where $H_{w2}$ acts.

For all simulations we choose $M = 453$ and $L = 300$ to make sure that there are no finite size effects 
associated with interfacial fluctuations. Also, in order to make it easier for
the reader to establish the connection to fluid droplets, we will describe all 
our results in terms of local densities defined via $\rho(i,j)=(\langle S(i,j) \rangle +1 )/2$.
Monte Carlo simulations were carried out with standard single spin-flip
algorithms \cite{41h}, which as is well known \cite{41h}, can be
interpreted as a simulation of a stochastic process, where (in a lattice gas
terminology) particles are randomly adsorbed at, or desorbed from, the sites
of the lattice; the rates of these processes satisfy the detailed balance 
condition with the Hamiltonian, Equation (\ref{eqh1}). Note that the density in the
considered lattice gas is not a conserved variable, of course; hence the
physical situation that is simulated is a two-dimensional substrate in
equilibrium with an ideal gas reservoir, at the specified temperature 
and chemical potential, corresponding to the chosen bulk magnetic field.
Of course, the Monte Carlo process has no intrinsic time units for the rates 
of Monte Carlo moves: so time is measured in units of "attempted Monte Carlo
steps per spin [MCS]" rather than any physical time units.
Simulations are started with an initial condition where all spins are taken 
as $S(i,j)=-1$, i.e.  $\rho(i,j) = 0$ in accord with the nonwet ground state 
of the system. Then, the system is equilibrated up to  $5\times 10^{6}$ 
MCS by taking $H_b = 0$, and subsequently is ``quenched'' to 
$H_b > 0$ in order to observe nucleation. Runs are performed for additional $2\times 10^{7}$ 
MCS, and averages are taken after disregarding $1\times
10^{7}$ MCS. 

\section{Wall Attached Precursor Droplets in Metastable Equilibrium.}

In our system, at least in the limit $M \rightarrow \infty$, $L \rightarrow \infty$,
the vapor phase is the true equilibrium phase only for $H_b \leq 0$. But when we consider 
a situation where equilibrium in the bulk has been established for $H_b = 0$,
and at the time of the Monte Carlo sampling that we take as the origin of time ($\tau = 0$) the field
$H_b$ is instantaneously switched to a small positive value, the vapor phase in the bulk 
may reach a state of metastable equilibrium, with a "lifetime" larger than the
observation time $\tau_{obs}$ of the simulation. This metastability can be understood 
qualitatively already by the classical theory of heterogeneous nucleation due to 
Volmer and Turnbull \cite{7hb,8h,9h,10h}, see the Appendix A.

In this region where the time $\tau_{N}$ to observe nucleation satisfies the condition 
$\tau_N \gg \tau_{obs}$, one may observe the formation of wall-attached precursor droplets
in the part of the sample where the surface field $H_{w3}$ acts, Figure \ref{fig1}.
Note that we restrict attention to temperatures $T$ distinctly lower than
$T_{w}$, i.e. when partial wetting of the wall occurs. To avoid confusion, we stress that
we denote the boundaries at $i = 1$ and $i = L$ in our system as "walls" although they are
one-dimensional lines only, Figure 1. If we would choose $T > T_{w}$, i.e. the case of 
complete wetting of the wall for $b \rightarrow \infty$, there would no longer   
occur any nucleation barrier, the liquid wetting layer that occurs then at the wall 
already for $H_b = 0$ would immediately grow by increasing its thickness as soon as  $H_b > 0$.   
For finite $b$ the interface of the liquid droplet still is pinned to the walls
near the sites where the boundary field changes from $H_{w3}$ to $H_{w2}$.
In our previous work \cite{53h} we have verified a prediction of 
Jakubczyk et al. \cite{55h,56h} for the case $H_b = 0$ and $T > T_w$,
based on the solid-on-solid (SOS) model in terms
of the interface Hamiltonian approach for the excess density $\Delta \rho$
due to the droplet, given by 

\begin{equation} \label{eqh2}
\Delta \rho=\Big(\rho^{\rm coex}_\ell - \rho^{\rm coex}_v \Big) \frac{1}{4} b^{3/2} \sqrt{\pi/2 \Sigma (T)} \quad , 
\quad b \rightarrow \infty \quad .
\end{equation}

\noindent In Equation (\ref{eqh2}), $\rho^{\rm coex}_\ell$ and $\rho^{\rm coex}_v$ are the coexisting
liquid and vapor densities in the bulk, respectively. Note that
$\rho^{\rm coex}_\ell - \rho^{\rm coex}_v = m^{\rm coex}$, the spontaneous magnetization of the
Ising model \cite{47h}.
Also, $ \Sigma (T)$ is the interfacial 
stiffness of the Ising model \cite{57h}.  Equation (\ref{eqh2}) implies that the average droplet shape
is a semi-ellipse, with small axis proportional to $b^{1/2}$ for $b \rightarrow \infty$. For large $b$  even
this small axis can easily exceed the critical droplet radius $R^{*}$ for homogeneous nucleation in 
the bulk (see Appendix A), and thus it is plausible that for
large but finite $b$ the nucleation barrier will be very 
small.  Although a study of the phase transformation for  $T > T_{w}$ and not so large $b$ may be 
interesting in its own right, we here fixed the attention to  $T < T_{w}$, where the linear dimension
in the $y-$direction of
the wall-attached droplet for $H_b = 0$ remains finite for $b \rightarrow \infty$. We have also 
shown \cite{53h} that the density in the mid-point $j_{half} = (M+1)/2$ of the inhomogeneity of the 
boundary is compatible with an exponential decay with the distance $y$ from the wall ($y$ is only 
defined at the discrete lattice indices $i$, of course),   

\begin{equation} \label{eqh3}
\rho (i, j_{half}) = A_0(t) \exp (-y/\xi_\bot (b,t)) , \quad H_b = 0 \quad .
\end{equation}

\noindent {\bf Note that in the context of various theoretical concepts on interfaces
  in this constrained geometry the use of a continuum description (in terms of coordinates
  $x$, $y$) is mandatory; but for a precise characterization of the simulation
  setup on the lattice, discrete indices of the lattice points ($j$ in $x-$direction,
  $i$ in $y-$direction) have also to be used.}
Here, $A_0(t)$ is an amplitude factor, $t = T/T_{cb}$ must be less than
$t_w = T_w/T_{cb}$, and the decay length  $\xi_\bot (b,t)$ converges to the standard transverse
correlation length $\xi_\bot (t)$ familiar from the theory of critical wetting \cite{49h} 
in $d = 2$ dimensions, with $\xi_\bot (t) \propto (t_w - t)^{-1}$. In order to avoid 
critical fluctuations associated with the second-order wetting transition, we consider
here only temperatures distinctly lower than $t_w$, namely $t = 0.30, 0.35$, and $0.40$   
(recall that $t_w = 0.4866$ for our choice of $h_{w3}$ \cite{48h,49h}).

\begin{figure}
\includegraphics[width=6cm,clip]{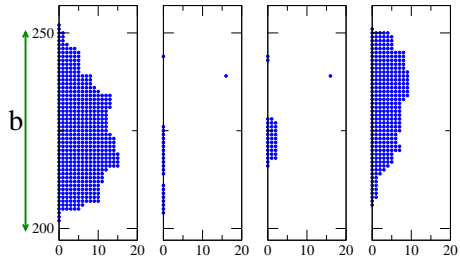}
\caption{\label{fig2} Snapshot configurations of the heterogeneous nucleation 
  of a droplet in an external field ($H_b$). Note that unlike Figure \ref{fig1},
  the boundaries are oriented along the ordinate direction, and the
  $y-$coordinate along the abscissa. Data obtained for $t = 0.40$,
and $b = 51$ (note the double arrow in the left-hand side of the figure
indicating the length $b$ of the heterogeneity). Each snapshot corresponds to different
realizations obtained with the same parameters but using different seeds in the 
random number generator. 
The simulated systems have size $L = 300$, and $M = 453$, while the snapshots
only show the central part, horizontal (vertical) length $20$ lattice units ($60$ lattice units), 
where the droplets develop. Note that the same length scale has been used for
both sides of the panels in order to display the actual shape of the droplets.
The system was equilbrated at $H_b = 0$ 
during $5\times 10^{6}$ MCS, subsequently "quenched" to $H_b = 0.030$, {\bf
  and all the snapshots are obtained for $\tau = 2\times 10^{7}$ MCS.} 
} 
\end{figure}

\begin{figure}[ht]
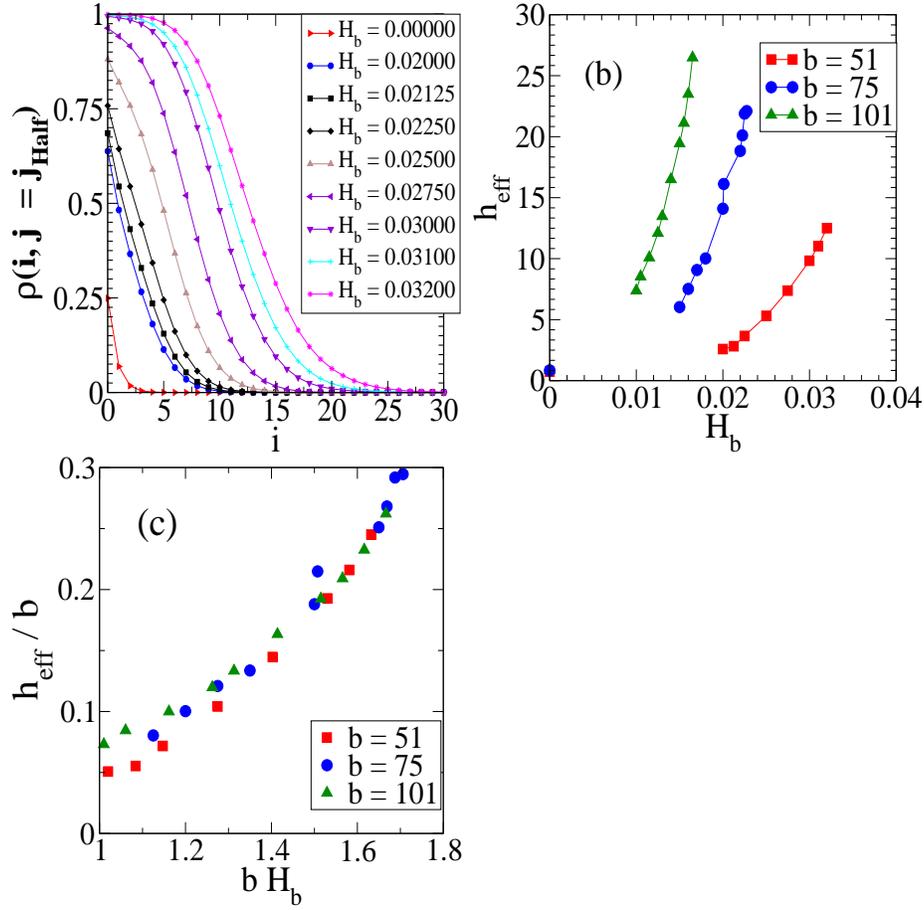

\includegraphics[width=6cm,height=6cm,clip]{Figure3a.eps}
\includegraphics[width=6cm,height=6cm,clip]{Figure3b.eps}
\includegraphics[width=6cm,height=6cm,clip]{Figure3c.eps}
  \caption{\label{fig3} 
a) Plots of the density profiles measured at the center of the 
sample ($\rho (i, j = j_{Half})$) {\it versus} the distance 
to the wall $y$, here $j_{Half} = (\frac{M}{2} + 1) = 227$. Of course, on the
lattice only discrete distances $y = 1, 2, 3,...$ are possible, so
the continuous curves are guides to the eye only.
Data corresponding to $t = 0.40$, $b = 51$, 
and different values of the bulk field, as indicated. 
b) Plot of the effective height ($h_{eff}$) of the droplet
as estimated by measuring the distance from the wall ($i = 1$)
to the inflection point of the density profile. Symbols at $H_b = 0$ 
correspond to the decay length  $\xi_\bot (b,t)$ obtained 
by fitting the profiles with the aid of Equation (\ref{eqh3}) \cite{53h},
namely  $\xi_\bot (101,0.40) = 0.8506$, $\xi_\bot (75,0.40) = 0.8379$, and 
$\xi_\bot (51,0.40) = 0.7338$.
Data obtained for $t = 0.40$ and three choices of $b$, as indicated.
The rightmost point for each choice of $b$ corresponds to the value
$H_b = H_{b}^{max}$, since for slightly larger choices of $H_b$ no
metastable droplets could be observed.
c) Scaling plot of $h_{eff}/b$ {it versus} $b H_b$ as obtained with the 
data already shown in panel b). 
}
\end{figure}

When we now consider metastable phases with small but nonzero $H_b > 0$, we find that
the actual droplet configurations are strongly fluctuating (Figure \ref{fig2}) even though the 
average profile $\rho(i, j = j_{half})$ is a smooth function (Figure \ref{fig3}(a)). We note that
for small enough $H_b$ (such as $H_b \leq 0.02$) the decay of the profile with the distance 
$y$ from the boundary still is compatible with Equation (\ref{eqh3}), and both the amplitude 
$A_0(t)$ and the decay length  $\xi_\bot (b,t)$ increases with $H_b$ only very slowly.
However, when we reach the apparent limit of metastability $H_{b}^{max}$, where for the 
chosen observation time $\tau_{obs} = 2 \times 10^{7}$ MCS nucleation becomes observable,
which in the case of Figure \ref{fig3}(a) is $H_{b}^{max} \approx 0.033$, the behavior changes:
In the regime $0.02 \leq H_{b} \leq 0.032$ the profiles change from a simple
exponential decay to sigmoidal, and $A_0(t)$  moves towards $\rho^{\rm coex}_\ell$, which
on the scale of Figure \ref{fig3}(a) is indistinguishable from unity, while $\rho^{\rm coex}_v$ is not 
distinguishable from zero here. Thus we have defined an effective droplet height $h_{eff}(t,H_b)$
by measuring the distance $y$ of the inflection point from the coordinate origin (Figure \ref{fig3}(b)). 
The increase of $h_{eff}(t,H_b)$ with $H_b$ is clearly faster than a straight line through 
the origin. An intriguing feature is the fact that the ratio $h_{eff}(t,H_b) / b$
seems to be a function of the scaling combination $b H_b$ only (Figure \ref{fig3}(c)).

\begin{figure}[ht]
\includegraphics[width=6cm,clip]{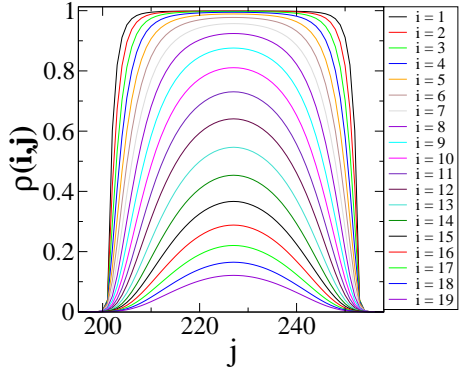}
  \caption{\label{fig4} 
Density profiles $\rho(i,j)$ measured in the $x-$direction parallel to the 
wall, and for different values of the distance $i-$ perpendicular to the wall, 
as indicated. Data correspond to $b = 51$, $t = 0.40$, and $H_b = 0.032$.
}
\end{figure}

In order to characterize the droplet shape more precisely, also density profiles 
$\rho(i, j)$ in the $x-$direction parallel to the boundary were taken (Figure \ref{fig4}). It is
clear that in the shown example the droplet has considerable extent in the $x-$direction,
comparable to $b$, as long as the distance $i$ from the boundary is clearly less than 
$h_{eff}$. Thus we have introduced a characteristic length of the wall-attached droplet in the 
$x-$direction, defining its "baseline" $b_{eff}$ as the area below the density profile for $i = 1$,
namely 
\begin{equation} \label{eqh4}
b_{eff} = \sum\limits_{j = (M-b)/2 + 1}^{(M +b)/2}  \rho(i = 1, j) \quad.
\end{equation}

\begin{figure}[ht]
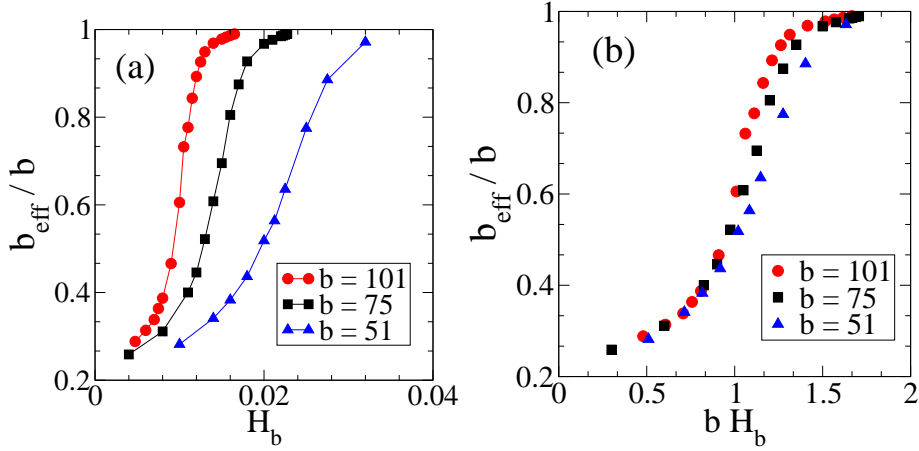

\includegraphics[width=6cm,clip]{Figure5a.eps}
\includegraphics[width=6cm,clip]{Figure5b.eps}
  \caption{\label{fig5} a) Plots of the normalized effective length of the droplet 
in contact with the wall (i.e. the ``baseline'') ($b_{eff}/b$) {\it versus}
$H_b$, as measured for the regime of wall-attached metastable droplets (precursor of the
actual nucleation that takes place for $H_b > H_b^{max}$. Results
obtained for $t = 0.40$, and different values
of $b$, as indicated. 
b) Scaling plot $b_{eff}/b$ {\it versus} $b H_b$, of the data already shown in panel a). 
}
\end{figure}

\noindent Figure \ref{fig5}(a) shows a plot of $b_{eff}/b$ for three choices of $b$ as a function of
$H_b$. One can see that for  $H_b \ll H_{b}^{max}$  $b_{eff}/b$ is small, but saturates
at unity when $H_b$ approaches $H_{b}^{max}$. Again the data for the three choices of 
$b$ almost superimpose on a master curve when one plots $b_{eff}/b$  as a function 
of the product $bH_b$ (Figure \ref{fig5}(b)); we shall discuss this scaling
behavior of the characteristic
lengths $h_{eff}$ and $b_{eff}$ below.

\begin{figure}[ht]
\includegraphics[width=6cm,clip]{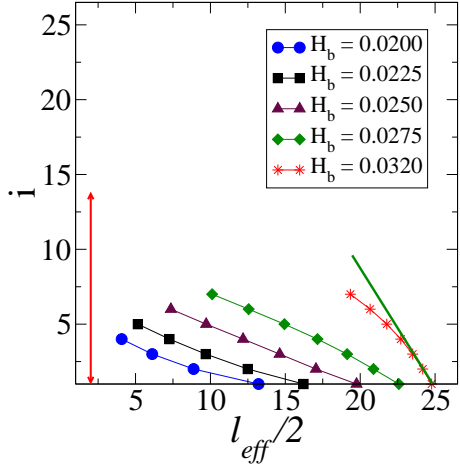}
  \caption{\label{fig6} 
Plot of the effective (half) length of the droplets ($l_{eff}/2$) as obtained from the 
integration of the density profiles as shown in figure \ref{fig4} (horizontal
axis), {\it versus} the 
distance to the wall where the inhomogeneity is placed (vertical axis). 
Data obtained for $t = 0.40$,
$b = 51$, and different values of the bulk field, as indicated.   
The double arrow at the left-hand side of the figure shows the effective height 
of the droplet ($h_{eff}$) as measured for $H_b = 0.032$ (see figure \ref{fig3}(b).)
The full straight line at the right-hand side of the figure shows the 
asymptotic slope of the droplet contour that is used to determine the effective contact angle
$\theta_{eff}$ (see also figure \ref{fig7}).
}
\end{figure}

\begin{figure}[ht]
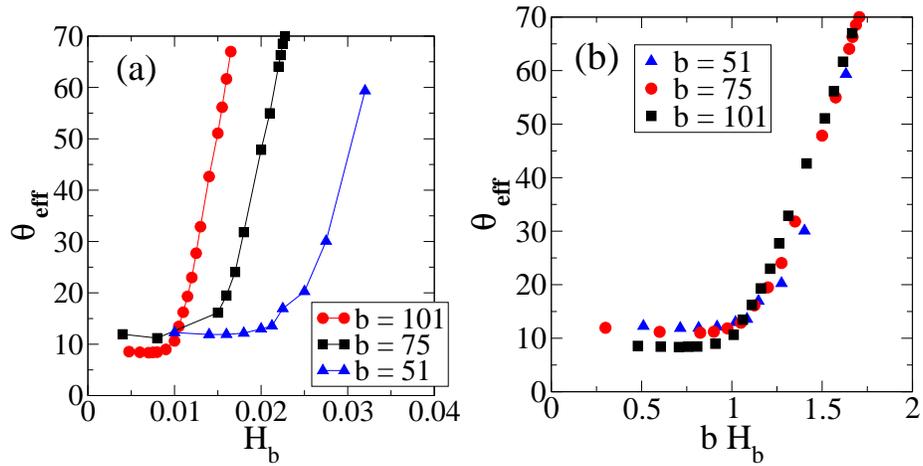

\includegraphics[width=6cm,clip]{Figure7a.eps}
\includegraphics[width=6cm,clip]{Figure7b.eps}
  \caption{\label{fig7}
a) Plots of the effective contact angle of the droplets ($\theta_{eff}$) {\it versus}
$H_b$, as measured for the regime of metastable wall-attached droplets
(precursor of the actual nucleation)
that takes place for $H_b <  H_b^{max}$, where $H_b^{max}$ is the value
  of the field where the transition to the liquid phase covering the
  whole sample is observed in the simulations.
  Results obtained for $t = 0.40$, and different values
of $b$, as indicated. 
b) Scaling plot $\theta_{eff}$ {\it versus} $b H_b$, of the data already shown in panel a). 
}
\end{figure}

When the ratio  $b_{eff}/b$ is no longer small, one can still use density profiles such as
those shown in Figure \ref{fig4} to evaluate the effective length of the droplet parallel
to the boundary but evaluated at a distance $i > 1$ from the wall (e.g. by performing 
the corresponding summation of Equation (\ref{eqh4})) and characterize the average 
shape of the droplet (note that this procedure is equivalent to the construction of contours 
of constant density), see Figure \ref{fig6}.  The slope of these contours can be used to extract estimates
for the effective contact angle $\theta_{eff}(H_b)$ from the part of the contours at 
small $i$ values, where these contours representing the coarse grained interface positions
reach the wall. Figure \ref{fig6} exploits this idea for the case $t = 0.40$, $b = 51$, and several
choices of $H_b$. By fitting straight lines to these contours in the region close to the wall
(i.e. for $i = 1, 2, 3$), estimates of $\theta_{eff}(H_b)$ can be extracted (Figure \ref{fig7}(a)). It is 
seen that for choices of $H_b$ for which $b_{eff} (H_b)$ is distinctly smaller than $b$,     
$\theta_{eff}(H_b)$ is essentially independent of $H_b$, and of the order of  
$\theta_{eff} \approx 10 \pm 2^{\circ}$ in this case. However, when $b_{eff} (H_b)$ starts to saturate
at $b$, the ratio $h_{eff} / b$ as well as the effective contact angle  $\theta_{eff}$ both start
to increase rather distinctly. As Figure \ref{fig7}(b) demonstrates, this increase of $\theta_{eff}$ 
starts at $b H_b \approx 1$. Metastable precursor droplets are found up to angles of 
$\theta_{eff}^{max} \approx 70^{\circ}$ when $H_b$ reaches $H_{b}^{max}$. 

\begin{figure}[ht]
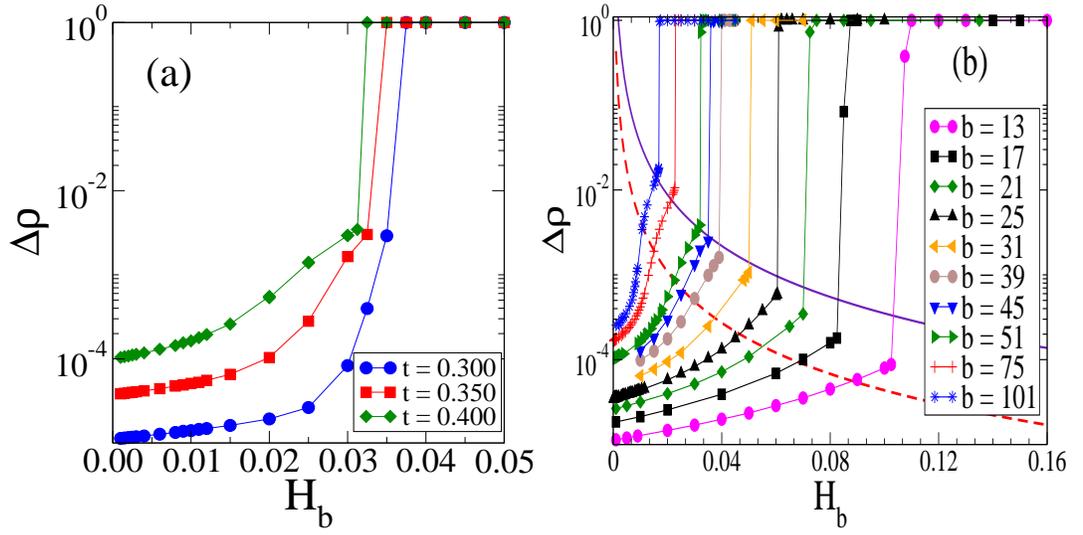

\includegraphics[width=7cm,height=7cm,clip]{Figure8a.eps}
\includegraphics[width=7cm,height=7cm,clip]{Figure8b.eps}
\caption{\label{fig8} a) Log-linear plot of the total density excess $\Delta \rho$
plotted {\it versus} $H_b$ for three 
choices of $t$, as indicated. Data obtained for $b = 51$. 
b) Log-linear plot of $\Delta \rho$ {\it versus} $H_b$ obtained at
temperature $t = 0.40$, and different values of $b$, as indicated.
The dashed (full) curve is a plot of $R^{*^{2}} \pi f_{VT} /LM$ (see Equation (\ref{eqh5}) noting
$\Delta \rho = \Delta m/(LM)$) {\it versus } 
$H_b$ as obtained by assuming 
$\theta_c = 34.155^{\circ}$ ($\theta_c = \theta_{eff}^{max} = 70^{\circ}$) in the evaluation of $f_{VT}$.   
More details on the text. 
}
\end{figure}

A related conclusion can be drawn with even less ambiguity, since it does not require 
an analysis of the shape of the precursor droplet, when we simply record the excess density
$\Delta \rho$ in the system due to the droplet (Figure \ref{fig8}). In fact, Figure \ref{fig8}(a)
shows plots of $\Delta \rho$ {\it versus} $H_b$ at three choices of $t$ for $b = 51$,
while Figure \ref{fig8}(b) shows plots of $\Delta \rho$ {\it versus} $H_b$ for $t = 0.40$ and
various choices of $b$, as indicated. Note that in both cases the states where the 
transition to the liquid in the simulated system has occurred are included, namely 
all the data points with  $\Delta \rho = 1$ implies that the whole simulation box 
is filled uniformly by a liquid. The excess density of the precursor droplet in these
plots is normalized by dividing the "excess mass" $\Delta m$ contained in the precursor 
droplet by the total number of lattice sites, $N = L \times M = 300 \times 453 = 135900$.
Recall that the excess density is the difference between the
density measured in the presence of the inhomogeneity in the wall and that obtained for an homogeneous wall.
Figure \ref{fig8}(a) implies that the value of $\Delta m$  where the precursor droplet is large
enough to trigger nucleation of the liquid phase in the system depends only weakly on temperature;
it is $\Delta m \approx 400$ for $b = 51$, and the choice $H_{w2} = 0.90$ that was made here.  
But again there is a pronounced dependence of $\Delta \rho$ (or $\Delta m$, respectively) 
at the transition on the choice of $b$ (Figure \ref{fig8}(b)). 

At this point it is interesting to make contact with the classical theory of heterogeneous
nucleation as formulated in $d = 3$ dimensions by Volmer and Turnbull \cite{7hb,8h,9h,10h}.
A simple adaptation of this theory to $d = 2$ (see Appendix A) implies that a critical droplet 
on a homogeneous substrate with contact angle  $\theta_{c}$ involves an excess density mass
given by
\begin{equation} \label{eqh5}
\Delta m = \Big(\rho^{\rm coex}_\ell - \rho^{\rm coex}_v \Big) (R^{*})^{2} \pi f_{VT}(\theta_{c})  \quad .
\end{equation}
\noindent Here, $R^{*}$ is the critical radius of homogeneous nucleation, which in the classical
theory simply is 
\begin{equation} \label{eqh6}
R^{*} = \frac{f_{int}}{2 ( \rho^{\rm coex}_\ell - \rho^{\rm coex}_v ) H_b} \quad ,
\end{equation} 
\noindent with $f_{int}$ being the interfacial tension between coexisting bulk phases
separated by a flat interface. As discussed in the Appendix A, Equations (\ref{eqh5}, {\ref{eqh6})
ignore the anisotropy of the interfacial tension, assuming a circular droplet in the 
bulk, and a circle cut shape of the droplet at the boundary, where the coarse-grained interface 
makes an angle $\theta_{c}$. The Volmer-Turnbull function $f_{VT}(\theta_{c})$ measures the reduction
of the droplet area of the circle cut relative to the full circle, and in this approximation (see Appendix A)
is given by 
\begin{equation} \label{eqh7}
f_{VT}(\theta) = \frac{1}{\pi} \Big( \theta - \frac{sin(2\theta)}{2} \Big) \quad .
\end{equation}
    
The upper curve in Figure \ref{fig8}(b) shows $\Delta \rho = \Delta m /LM$ as
obtained according to 
Equations (\ref{eqh5}-\ref{eqh7}) by using the "observed" contact angle
$\theta_{c} = \theta_{eff}^{max} \approx 70^{\circ}$ just at the
transition to the liquid phase. However, by using a theoretical estimate of
the contact angle in thermal equilibrium obtained by means of a SOS  approximation $\theta_{c}$ 
due to Abraham et al. \cite{50h} would imply $\theta_{c} \approx 34^{\circ}$, predicting hence
distinctly smaller critical droplets (Figure \ref{fig8}(b)). However, for both choices of  
$\theta_{c}$ our results (Figures \ref{fig5}, \ref{fig7}) imply that $b_{eff}$ is of the same order as $b$
when nucleation occurs. Since the baseline length $b_{drop}^{*}$ of the circle cut critical droplet 
satisfies the geometrical relationship
\begin{equation} \label{eqh8}
b_{drop}^{*} = 2 R^{*} sin(\theta_c)  \quad,
\end{equation} 
\noindent we can eliminate $R^{*}$ in Equations (\ref{eqh5}, \ref{eqh6}) in
favor of $ b_{drop}^{*} = b$, and in this way the two theoretical curves in Figure \ref{fig8}(b) were
obtained. Figure \ref{fig9}(a) shows then the scaling plot of  $\Delta \rho /b^{2}$ {\it versus}
$b H_b$, validating the idea that  $\Delta m$ scales like $b^{2}$ and is a function of 
the product $b H_b$, as $b_{eff}$ and $h_{eff}$ (Figures \ref{fig3}(c) and \ref{fig5}(b)).  
Both figures  \ref{fig7} and \ref{fig8} show that two regimes exist: for $bH_b \lesssim 1$,
$\theta_{eff}$ is very small, $l_{eff} < b$ (Figure \ref{fig8}), $\Delta \rho$ increases
slowly with $bH_b$, and the droplet density profile (Figure \ref{fig3}(a) ) decays with the
distance from the inhomogeneity like an exponential. All these properties change for
$bH_b \gtrsim 1$. Note that  \ref{fig9}(a) includes both the regime
$bH_b \lesssim 1$, where scaling is not expected hold, as well as the true
scaling regime $bH_b \gtrsim 1$. Of course, for $H_b = 0$ in the nonwet regime the
excess density $\Delta \rho$ only scales like $\Delta \rho \propto b$; thus in the
regime of small $bH_b$ a crossover from $\Delta \rho \propto b$ to $\Delta \rho \propto b^{2}$
with increasing $bH_b$ must occur.

\begin{figure}[ht]
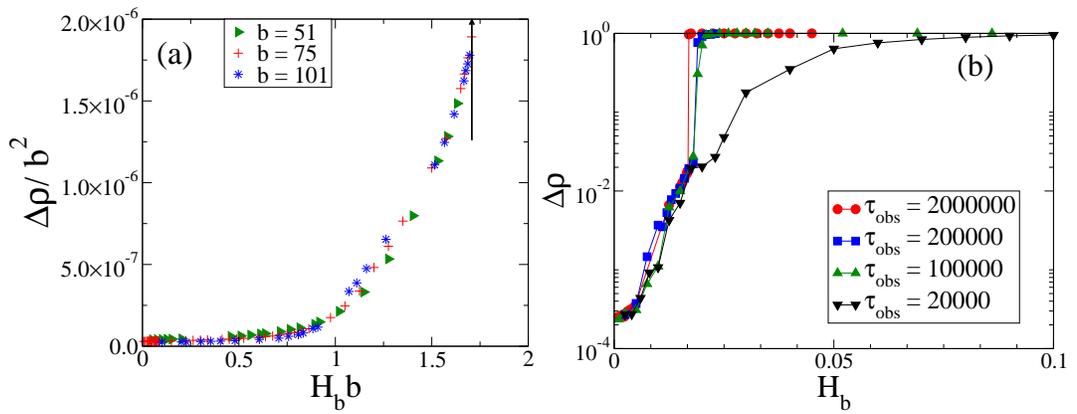

\includegraphics[width=7cm,clip]{Figure9a.eps}
\includegraphics[width=7cm,clip]{Figure9b.eps}
\caption{\label{fig9} a)  Scaling plot of the total density excess, $\Delta \rho /b^{2}$  {\it versus} $b H_b$,
according to Equation (\ref{eqhqyq}). Data taken from figure \ref{fig8}(b).  
The vertical arrow at the right-hand side of the figure indicates the asymptotic 
value {\bf $b H_b^{max} \simeq 1.71$} for the observation of nucleation, which yields 
{\bf $\theta_c^{max} \approx 70^{°}$} as it follows from Equations (\ref{eqh6}) and (\ref{eqh8}). 
b) Log-linear plot of the total density excess, $\Delta \rho$  plotted
{\it versus}  $H_b$, for the case $t = 0.40$, $b = 51$, and 4 choices of the
observation time $\tau_{obs}$ of the metastable state, as indicated.
}
\end{figure}

In order to give a physical interpretation of the simple scaling of all quantities ($b_{eff}$,
$h_{eff}$, and $\Delta \rho$) with the product $b H_b$, we recall the description of droplets at
chemical inhomogeneous substrates in $d = 2$ dimensions in terms of the interface Hamiltonian
proposed by Jakubczyk et al.\cite{55h,56h}. In this description in the spirit of a Solid on
Solid (SOS) model, the problem is described by a one-dimensional degree of freedom, namely 
the distance  $y = \ell (x)$ of the (locally sharp) interface from the boundary at 
$y = 0$ (involving a continuum approximation). So the effective (coarse-grained) Hamiltonian
is, absorbing a factor $\frac{1}{k_{B}T}$ here, 
\begin{equation} \label{eqhxx}
\mathcal{H} \Big[\ell(x) \Big] =\int\limits^{+M/2}_{-M/2}\, dx \Big[\frac{\Sigma (T)}{2} \Big(\frac{d \ell}{dx}\Big)^2 + V(x, \ell) \Big] ,
\end{equation} 
\noindent where both fluctuations in the bulk and overhangs of the interface are neglected,
$V(x, \ell)$ being the effective potential acting on the interface. Recall
  that in the SOS treatment the interfacial stiffness $\Sigma (T)$ \cite{57h} of the
  one dimensional interface is considered, instead of the actual interfacial
  tension $f_{int}$ of the Ising model \cite{46h}. For the considered situation,
for $|x| > b/2$ the boundary at $y = 0$ strongly favors the vapor phase, so we have
essentially  $\ell (x) = 0$ there, as one can verify from Figure \ref{fig4}. So, Equation (\ref{eqhxx}) 
can be reduced to   
\begin{equation} \label{eqhyy}
\mathcal{H} \Big[\ell(x) \Big] =- \int\limits^{+b/2}_{-b/2}\, dx \Big[\frac{\Sigma (T)}{2} \Big(\frac{d \ell}{dx}\Big)^2 + V(x, \ell) \Big] .
\end{equation}
\noindent Note that Equations (\ref{eqhxx}, \ref{eqhyy}) also assume $|\frac{d \ell}{dx}| \ll 1$
everywhere, an assumption that is somewhat questionable in view of the actual snapshots of the
interfacial configurations (Figure \ref{fig2}), at least near $x = \pm \frac{b}{2}$;
however since no actual calculations on the basis of Equations (\ref{eqhxx}, \ref{eqhyy}) are done here,
this problem does not matter.

Now the effective potential  $V(x, \ell)$ can be written as 
\begin{equation} \label{eqhzz}
V(x, \ell) = \Big[V_{0}(x, \ell) - \ell \Big(\rho^{\rm coex}_\ell - \rho^{\rm coex}_v \Big) H_b\Big] /k_BT \quad , 
\end{equation}
\noindent where $V_{0}(x, \ell)$ is the potential binding the interface to the wall for $H_b = 0$.
Only this latter case has been considered by Jakubczyk et al.\cite{55h,56h}. Applying a field 
$H_b > 0$ favors the liquid phase, and thus the potential decreases proportional to $\ell H_b$.

Now the key observation is that the dependence on $b$ is elucidated when we rescale all 
distances by $b$, namely 
\begin{equation} \label{eqhww}
x = b x ' \quad,   \ell = b \ell ' \quad ,
\end{equation}    
\noindent which yields $\mathcal{H} = b\mathcal{H}'$ with
\begin{equation} \label{eqhss}
\mathcal{H}' \Big[\ell'(x')  \Big] =- \int\limits^{+1}_{-1}\, dx' \Big[\frac{\Sigma (T)}{2} \Big(\frac{d \ell'}{dx'}\Big)^2 + V(x', \ell') \Big] ,
\end{equation}
\noindent with 
\begin{equation} \label{eqhqq}
V(x', \ell') = V_{0}(x', \ell') - \ell' \Big(\rho^{\rm coex}_\ell - \rho^{\rm coex}_v \Big) b H_b \quad . 
\end{equation}
\noindent Now the partition function needs to be evaluated as a path integral,
\begin{equation} \label{eqhrr}
\mathcal{Z} =  \int \mathcal{D}\ell' exp\Big(-b\mathcal{H}' \Big)    \quad, 
\end{equation}
\noindent and from Equations (\ref{eqhss})-(\ref{eqhqq}) we conclude that the boundary excess
free energy $\Delta F = -k_{B} T ln (\mathcal{Z})$ due to the droplet depends on the variables $b, t$,
and $H_b$ in the following scaled form, $f(t, bH_b)$ being the free energy density per length unit along 
the boundary 
\begin{equation} \label{eqhquq}
\Delta F =  b f(t, bH_b)  \quad .
\end{equation}

The excess density due to the droplet is obtained from Equation (\ref{eqhquq}) via a 
derivative with respect to $H_b$, i.e.
\begin{equation} \label{eqhqyq}
\Delta \rho =  b^{2} \tilde M(t, b H_b)  \quad ,
\end{equation}
\noindent where $\tilde M$ is the resulting scaling function of the excess mass. Equation (\ref{eqhqyq})
hence justifies the choice of scaling variables for Figure \ref{fig9}(a).

This scaling property is subtle, of course, due to the requirement of metastable equilibrium:
it is implied also by Equation(\ref{eqhzz}), that there cannot be for $H_b > 0$ a true equilibrium 
at any finite value of $\ell$, so Equation (\ref{eqhrr}) makes sense only for a suitably constrained 
partition function.

So the droplets studied so far can only be found in a suitable "window" of observation times
$\tau_{obs}$. In fact, $\tau_{obs}$ must be large enough to allow that the wall-attached droplet
reaches local equilibrium in spite of the slow and sluggish  fluctuations of the interface 
configuration (Figure \ref{fig2}). But at the same time, $\tau_{obs}$ must be small enough that
nucleation events (where the droplet grows fast to the full size of the system, see Section V)
still are negligible.

This consideration is exemplified in Figure \ref{fig9}(b): here a log-log
plot of the excess density in the system {\it versus} $H_b$ is shown, for four
choices of  $\tau_{obs}$. For very small $H_b$, such as $H_b = 0.0025$ the
dependence on  $\tau_{obs}$ is negligible since the time  $\tau_{N}$ needed to
nucleate is astronomically large, and the wall attached droplet is very tightly
bound to the wall (cf. Figure \ref{fig3}(a)), so it is rather easily
equilibrated.  However, for  $H_b = 0.01$ we see that data for
 $\tau_{obs} = 2 \times 10^{5}$ and $\tau_{obs} = 2 \times 10^{6}$ perfectly
agree, nucleation is not yet possible; but the result for 
$\tau_{obs} = 2 \times 10^{4}$ is clearly smaller, this observation time was 
insufficient to sample fluctuations such as those seen in Figure \ref{fig2} 
exhaustively. For  $H_b = 0.018$ however, there is also a systematic
difference between  $\tau_{obs} = 2 \times 10^{5}$ and $\tau_{obs} = 2 \times
10^{6}$: for the latter time, nucleation typically has occurred, while for the
former time, the metastable boundary-attached droplet still is visible.

Since Figure \ref{fig5} suggests that we can (for the choices of $b$ used
here) observe metastable boundary attached droplets up to $b_{eff} \approx b$,
we have also tested as a possible hypothesis that these boundary-attached
droplets with  $b_{eff} = b$ can be described by the Volmer-Turnbull theory of
heterogeneous nucleation (see the Appendix A). This theory assumes that the
critical droplet causing nucleation is a cut from a sphere (circle in our $d =
2$ case) with radius $R^{*}$ [Equation (\ref{eqh6})], the angle of the
sphere cut with the straight line representing the boundary being 
the contact angle $\theta_c$. Geometry then implies $b_{eff} = b$ as quoted in 
Equation (\ref{eqh8}), and combining Equations (\ref{eqh6}), (\ref{eqh8})
yields a relationship between $b$ and $H_{b}$
\begin{equation} \label{eqhaa}
b_{drop}^{*} = b = f_{int} sin(\theta_c) / \Big(\rho^{\rm coex}_\ell - \rho^{\rm coex}_v \Big)  H_{b}^{crit} \quad , 
\end{equation}
\noindent  where  $H_{b}^{crit}$ is the prediction of standard theory of
  heterogeneous nucleation for the
  critical field at the onset of nucleation of the liquid phase when
  the length of the baseline of the droplet is $b_{drop}^{*} = b$.
This result is plotted in Figure \ref{fig10}, using for $f_{int}$
the Onsager result \cite{46h} for the interface tension of a straight
interface oriented perpendicular to the lattice axis. For the contact angle    
$\theta_c$ we use results derived for Abraham et al. \cite{50h} in the SOS
approximation, namely 
\begin{equation} \label{eqang}
tan [\theta_c(T_w,T)] = sinh \Big[2 (K - K_w \Big] / \Big[ cosh(K) - cosh[2(K -
K_w)] \Big]  \quad ,
\end{equation}
\noindent where $K = J/k_B T$ and  $K_w = J/k_B T_w$, respectively.

\begin{figure}[ht]
\includegraphics[width=7cm,clip]{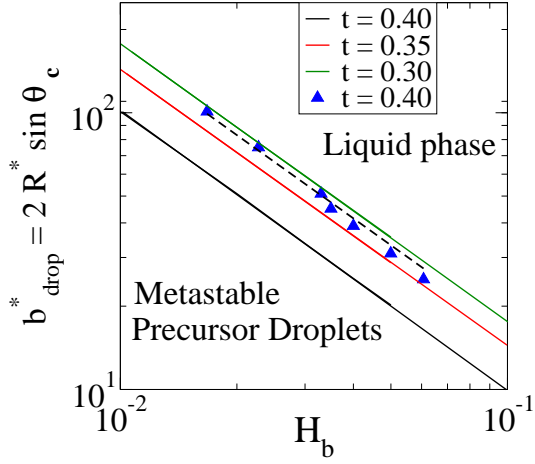}
\caption{\label{fig10}
Log-log plot of $b_{drop}^{*} = 2 R^{*} sin(\theta_c)$ {\it versus} $H_b$ 
corresponding to different temperatures as indicated.
Full lines correspond to the theoretical prediction  (Equation (\ref{eqhaa}))
that follows by considering the values of $\theta_{c}$ given
by a SOS calculation (Equation (\ref{eqang})). The straight lines have slope $-1$.
Numerical data corresponding to $ t = 0.40$ are shown by full triangles.
The dashed line is a fit of the numerical data, obtained by assuming a
slope $-1$, which yields  $\theta_{c}^{max} = \theta_{eff}^{max} =  67.5^{\circ}$
if Equation (\ref{eqhaa})
is invoked for the numerical simulation data.
At each temperature the full straight lines are the theoretical curves where
nucleation according to the Volmer-Turnbull theory is predicted to occur
when the baseline of the critical droplet equals $b$, so that they
are the boundary between the region where 
metastable precursor droplets are observed (lower left-hand side of the
panel), and the liquid phase (upper right-hand side of the panel). 
More details on the text. 
}
\end{figure}

For our choice of $H_{w3}$ implying $T_w = 0.4866 T_{cb}$, Equation
(\ref{eqang}) yields $\theta_c \approx 67,11^{\circ}$, $50,05^{\circ}$ and
$34,11^{\circ}$, for $t = T/T_{cb} = 0.30, 0.35$ and $0.40$, respectively. 
The resulting straight lines in the log-log plot 
for $b$ {\it versus} $H_{b}$ [Equation (\ref{eqhaa})] are compared 
to the estimates for the actual critical field $H_b^{max}$ where the 
onset of nucleation of the liquid phase has been observed
in the simulations [Figure \ref{fig10}]. 
For each temperature, these lines correspond to the theoretical 
conditions where nucleation on the inhomogeneity of length $b$ becomes possible.
The actual data points included in Figure \ref{fig10} separate the region of parameters where 
metastable precursor droplets are found, and the liquid phase that 
already takes the whole simulation box.   
It is seen that the actual critical fields, measured for $t = 0.40$ 
always are larger than the predictions
based on Equations (\ref{eqhaa}), (\ref{eqang}). So, if we would fit the numerical data
to equation (\ref{eqhaa}) we would obtain
the  "observed" contact angle $\theta_{c}^{max} = \theta_{eff}^{max} \approx 67.5^{\circ}$, in
excellent agreement with our previous estimations, e.g. showing that 
metastable precursor droplets are found up to angles of 
$\theta_{eff}^{max} \approx 70^{\circ}$ (Figure \ref{fig7}(b)), as well
from the scaling plot of Figure \ref{fig9} that also 
yields $\theta_{eff}^{max} \approx 70^{\circ}$}. The main reason for the difference between the
field $H_b^{crit}$, predicted by the standard theory for heterogeneous nucleation and
  defined from equation (\ref{eqhaa}) and calculated by using the contact angle $\theta_c$
obtained by means of the SOS approximation (equation (\ref{eqang})), and the actual critical
field found in the simulations $H_b^{max}$ is that in
the regime $H_b^{crit} < H_b < H_b^{max}$ the droplets nucleated
at the inhomogeneity are pinned, see Section IV, i.e., their baseline cannot grow
beyond $b$. However, it should be stressed that within this regime the
area of the droplets actually grows by simultaneously increasing the contact angle and
decreasing their radius. Only for $H_b > H_b^{max}$ droplets ``depin'' and further growth is possible,
with $b_{drop} > b$ and $\theta = \pi - \theta_c$, see also below.
These pinned droplets should not be mistaken for the droplets described by the
standard Volmer-Turnbull theory of heterogeneous nucleation, as discussed in 
the Appendix A.
In Section IV, we shall attempt a theoretical estimation of the field $H_b^{max}$. We also
note that for $H_b < H_b^{crit}$ only subcritical nuclei ($R < R^{*}$) can form
on the inhomogeneity, i.e. transient fluctuations occur whose average effect shows
up in the exponentially decaying density profiles for $H_b \leq 0.022$
in figure \ref{fig3}(a).

\begin{figure}[ht]
\includegraphics[width=14cm,clip]{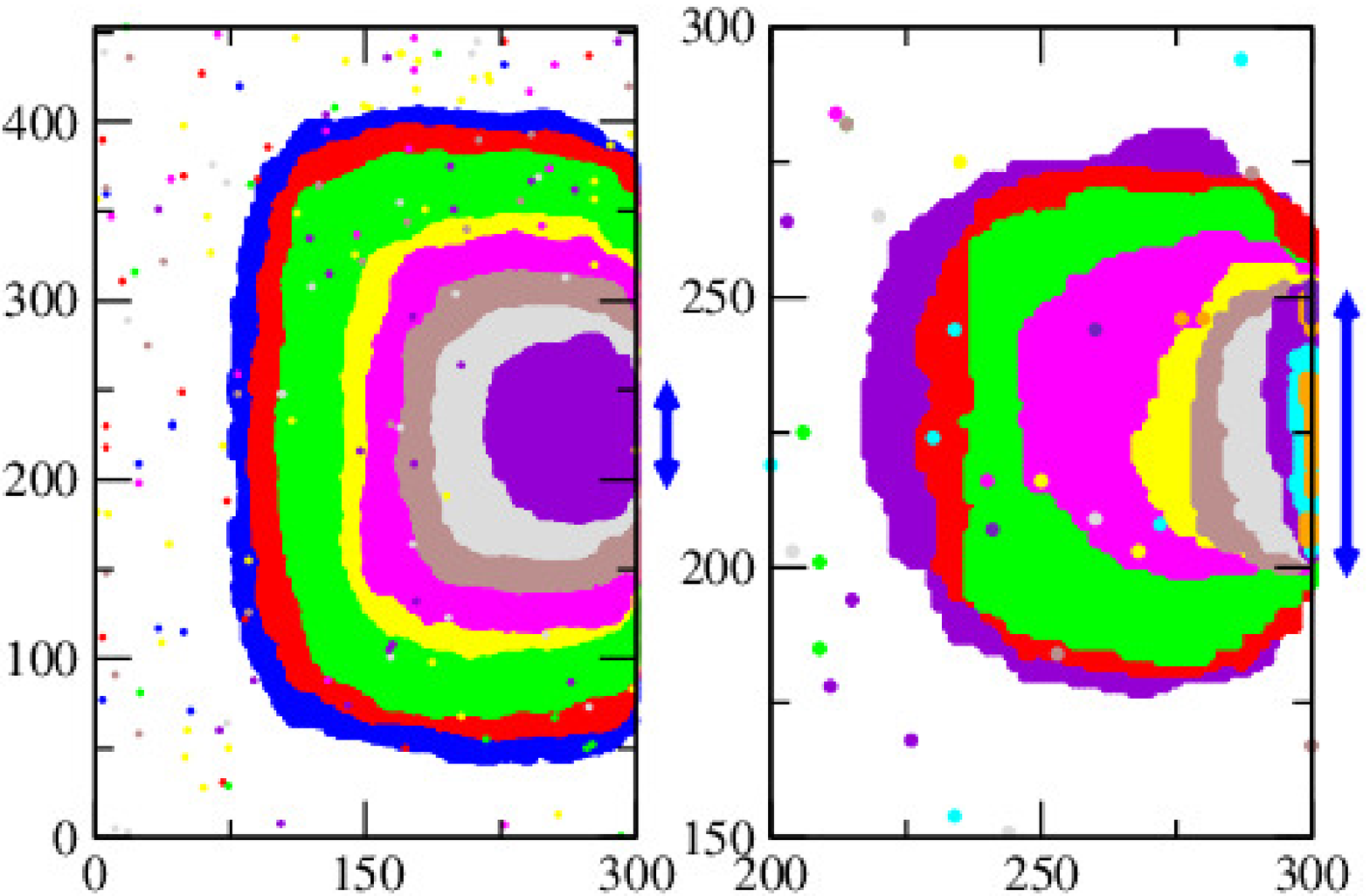}
\caption{\label{fig11} Snapshot configurations of the heterogeneous nucleation     
of a droplet in an external field ($H_b$). Data obtained for $t = 0.40$,
$b = 51$ (note the double arrow in the right-hand side of each panel
indicating the length of the heterogeneity), 
$L = 300$, and $M = 453$. The system was equilbrated at $H_b = 0$ 
during $5\times 10^{6}$ MCS, and subsequently "quenched" to $H_b = 0.034$.  The 
snapshot is recorded when the density of the nucleated droplet reaches the 
different values, which are identified by different colors. The
simulation times ($\tau$) in MCS, necessary to reach each density are also listed
between bracket. Then, going from high to low density the values and colors are:
Panel a) 0.556  [172094] (blue), 0.493 [171124] (red), 0.420 [169948] (green),
0.260 [167043] (yellow), 0.221 [166148] (magenta), 0.153 [164171] (brown),
0.100 [162312] (grey), and 0.052 [159035] (violet).
Panel b) 0.052 [159035] (i.e. last value and color from panel a), 0.044 [158518] (red),
0.037 [157165] (green), 0.022 [152678] (magenta), 0.010 [74795] (yellow),
0.0075 [31525] (brown), 0.0050 [12212] (grey), 0.0025 [1773] (violet), 0.0010 [932] (cyan),
0.0008 [845] (magenta), and 0.0006 [787] (orange).
Note the different scales for abscissa and ordinate in both panels;
while a) shows the full system, b) only shows a subpart of the system
that contains the wall inhomogeneity plus attached droplet.
More details in the text.
{\bf Warning. In the final editing process the ratio H/W = High/Width
of each panel must be kept at $H/W = 3/2$.}
}
\end{figure}

Furthermore, it is worth discussing that Equation
(\ref{eqhaa}) is not expected to be quantitatively accurate for several reasons: (i) The
interface tension for a straight interface $f_{int}$ is used here, neglecting
possible corrections due to the curvature of the droplet interface. (ii) Due
to the anisotropy of the interfacial tension of the Ising lattice model, the
actual shape of a large droplet of the liquid coexisting with surrounding
vapor is not a circle, it rather resembles a square with
rounded corners at low temperatures \cite{60h,61h} (see also the
largest droplet in Figure \ref{fig11}, left-hand side panel). For heterogeneous
nucleation, the droplet shape resulting from the appropriate Winterbottom
construction \cite{62h} is then nontrivial to find, and the
Volmer-Turnbull theory as presented in the Appendix A needs to be extended to
account for this anisotropy. For not very large droplets, also 
the ``point'' where the droplet interface meets the boundary can play a role,
modifying Equation(\ref{eqhaa}) further, in analogy with the effect of the
line tension of the sphere-cap shaped droplet
on the contact angle in $d = 3$ dimensions \cite{90}.
In view of all these shortcomings of the existing theories, a 
more quantitative analysis of our numerical
data for the boundary-attached droplets (Figures \ref{fig3} - \ref{fig9})
suffers from the incomplete knowledge of both the droplet shape and the equilibrium
contact angle. Nevertheless, we attempt a phenomenological analysis of pinned
droplets and their depinning in the next section.

\section{Droplets pinned at chemical inhomogeneities and their ``depinning transition''.}

When one deals with heterogeneous nucleation at homogeneous substrates, the nucleation
barrier $\Delta F_{het}^{*}$ (Equation (\ref{A13})) corresponding to the droplet having the
critical radius $R^{*}$ (equation (\ref{eqh6}) or (\ref{A13}), respectively) is all what matters:
when such a droplet (of circle cut shape, with contact angle $\theta_c$) corresponding
to the top of the free energy $\Delta F_{drop}(R)$ (equation (\ref{A12})) has been formed by a
(rare) statistical fluctuation, with $50 \%$ probability this drop will grow with time $\tau$
after the nucleation event. For small fields $H_b$ the growth velocity is small, and then
we have ``local equilibrium'' of the growing droplet at the contact line; this means, at
any instant of time growing droplets with $R > R^{*}$ are still described by equation (\ref{A12}),
and, in particular, their contact angle has the equilibrium value $\theta_c$.

However, this description cannot apply when we have a substrate with a chemical inhomogeneity
of extent $b$ (c.f. figure \ref{fig1}). We assume here conditions (corresponding to our actual
choice of the boundary fields $H_{w1}$, $H_{w2}$, and $H_{w3}$) where nucleation rates in the region
where wall fields $H_{w1}$, $H_{w2}$ act are negligibly small; so only nucleation within the region
of the chemical inhomogeneity needs to be considered, i.e. circle-cut shaped droplets with baseline
$b_{drop} = 2 R sin(\theta_c)$ (equation (\ref{A1})) smaller than $b$. Such droplets can grow
at constant contact angle with time only until $b_{drop} = b_{drop}^{*} = b$, and then
either get pinned and grow
in area and angle up to some nontrivial values, which we shall study 
in this section or they ``depin'' and grow with baseline $b_{drop} > b$, $b_{drop} = 2 R sin(\pi - \theta_c)$.

Thus we turn to an analysis of the regime where $b_{drop} \simeq b$.
In this regime we have to use Equation (\ref{A9})
for the area of the droplet, and hence write the free energy of the droplet as
\begin{equation} \label{A16}
\Delta F_{drop} = constant +  f_{int} 2 R \theta - 2 m_{coex} H_b R^{2}[\theta - \frac{1}{2} sin(2 \theta)] \quad ,
\end{equation} 
\noindent where the constant is fixed by the requirement that for $\theta = \theta_c$ and $b_{drop} = b$
the previous expression for $\Delta F_{drop}$ (equation (\ref{A12})) results,
i.e. $constant = -b f_{int} cos(\theta_c)$. Note
that now $R$ is not $R^{*}$ but rather $R = b/(2 sin(\theta))$ from geometry (see
Figure \ref{FigA2} in Appendix A).
Thus we obtain, eliminating $R$ in favor of $b/(2 sin(\theta))$,
\begin{equation} \label{A17}
\Delta F_{drop} /b f_{int} = -cos(\theta_c) + \frac{\theta}{sin(\theta)} - \frac{m_{coex} H_b b}{2 f_{int}} [\frac{\theta}{sin^{2}(\theta)} - \frac{cos(\theta)}{sin(\theta)}] \quad .
\end{equation} 

Now the angle $\theta$ is found from the condition
\begin{equation} \label{A18}
\frac{\partial}{\partial \theta} (\Delta F_{drop}/bf_{int}) = 0  \quad ,
\end{equation} 
\noindent which after simple algebra yields the minimum of the free energy for  
\begin{equation} \label{A19}
sin(\theta_{min}) = m_{coex} H_b b /f_{int}   \quad , \quad  \theta_c < \theta_{min} < \pi/2  \quad ,
\end{equation} 
\noindent while the angle $\theta_{max} = \pi - \theta_{min}$ also is a solution of equation (\ref{A18}),
but corresponds to the maximum of the free energy. Note that the condition $\theta_c < \theta_{min} $
has been added since equation (\ref{A17}) makes sense only for $\theta > \theta_{c}$. 

\begin{figure}[ht]
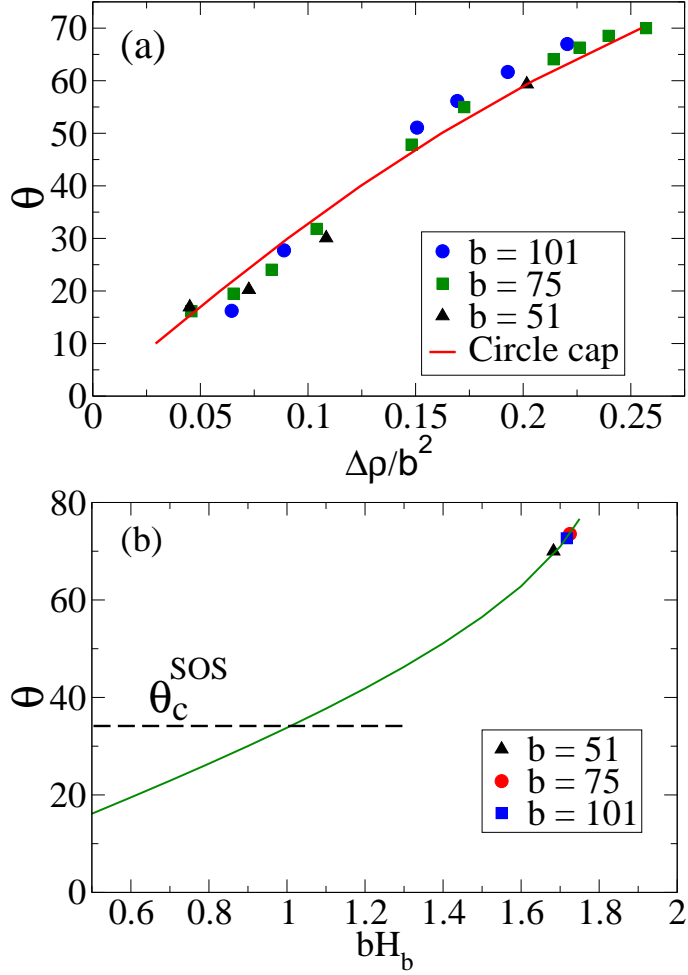

  \includegraphics[width=9cm,clip]{Figure12a.eps}
  \includegraphics[width=9cm,clip]{Figure12b.eps}
  \caption{\label{FigA4}
    a) Plots of $\theta$ {\it versus} $\Delta \rho /b^{2}$, comparing Equation (\ref{A9}) to the simulation
data, obtained for $t = 0.40$, in the regime $20^{\circ}  < \theta < 70^{\circ}$. 
b) The full line shows a plot of $\theta$ {\it versus} $b H_b$, according to equation (\ref{A19}) taken
$f_{int} = 1.7987$ for $t =  0.40$ \cite{46h}. The dashed 
horizontal line shows the estimate of $\theta_c$ from the SOS approximation Equation (\ref{eqang}).
Also, simulation results of $\theta_{eff}^{max}$ obtained for
different values of $b$ are shown by mean of symbols.
}
\end{figure}

Using this result it is straightforward to evaluate the area $A$ of the droplet and hence 
the excess density $\Delta \rho$ due to the droplet.  Figure \ref{FigA4} a) presents 
plots of $\theta$ {\it versus} $\Delta \rho /b^{2}$, comparing Equation (\ref{A9}) to the simulation
data in the regime $20^{\circ}  < \theta < 70^{\circ}$. The agreement
is reasonable, in particular since no adjustable
parameter whatsoever is present. Note that the knowledge of the contact angle $\theta_c$ is not needed
here (apart from defining the range on which this relationship should be used).

Figure \ref{FigA4} b) presents a plot of $\theta$ {\it versus} $b H_b$. The dashed 
horizontal line shows the estimate of $\theta_c$ from the SOS approximation Equation (\ref{eqang}). The 
actual variation of $\theta$ with $b H_b$ (Figure \ref{fig7}b ) should be compared to this 
figure only for $b H_b > 1$, since for small fields, where the excess mass due to the droplet is small,
the assumptions of the above quasi-macroscopic analysis clearly are inapplicable. At least, for   
$b H_b \ge 1.5$ the prediction is close to the observations from the simulations (Figure \ref{fig7} b),
furthermore the values of $\theta_{eff}^{max}$ obtained for the larger inhomogeneities,
i.e. $51 \leq b \leq 101$, are in full agreement with the theoretical result given
by equation (\ref{A19}) taken
$f_{int} = 1.7987$ for $t =  0.40$ \cite{46h} ($f_{int}$ is taken in units of $J$).    

\begin{figure}[ht]
\includegraphics[width=14cm,clip]{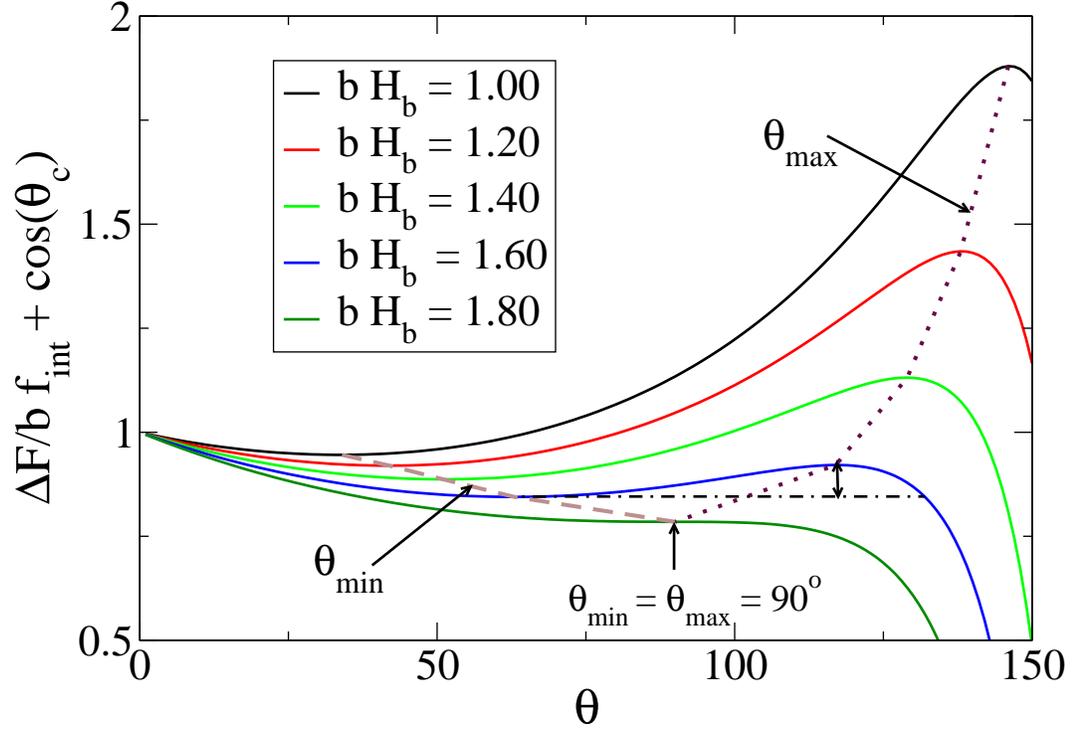}  
\caption{\label{fig13}
Plots of the free energy $\Delta F / b f_{int} + cos(\theta_c)$ given by equation (\ref{A17}) as function
of $\theta$ for various  values of the parameters $b H_b$, as indicated.
Plots obtained for $\theta_c = 34^{\circ}$ and $f_{int} = 1.7987$, which correspond
to $t = 0.4$.
The dashed (dotted) line shows the location of the minimum (maximum) of
equation (\ref{A17}), as given by equation (\ref{A19}). Note that
both lines merge at $\theta = \pi/2$ such that $\theta_{min} = \theta_{max}$.
The double arrow shows the free energy barrier $\Delta F_2 / b f_{int}$ given
by equation(\ref{A20a}), for the choice $b H_b = 1.6$.
}
\end{figure}
It really is illuminating to plot the free energy given by equation (\ref{A17}) as function
of $\theta$ for various values of $b H_b$, as shown in figure \ref{fig13}. In fact, both the
minimum and the maximum of the free energy can clearly be observed. Furthermore, the
free energy difference between the maximum and the minimum ($ \Delta F_2 / b f_{int}$)
monotonically decreases when $b H_b$ increases, and finally vanish
where $\theta_{min}$ and $\theta_{max}$ merge at $\theta_{min} = \theta_{max} = \frac{\pi}{2}$.
A more quantitative evaluation of the free energy barrier can be performed  
by reinserting Equation (\ref{A19}) in the free energy  $\Delta F_{drop}$ (equation (\ref{A17})),
obtaining the following relationships for the minimum
\begin{equation} \label{A20}
\Delta F_{drop}^{min} /b f_{int} = -cos(\theta_c) + \frac{1}{2}[\frac{\theta_{min}}{sin(\theta_{min})} + cos(\theta_{min})] \quad ,
\end{equation}
\noindent and the maximum
\begin{equation} \label{A20b}
\Delta F_{drop}^{max} /b f_{int} = -cos(\theta_c) + \frac{1}{2}[\frac{\pi - \theta_{min}}{sin(\theta_{min})} - cos(\theta_{min})] \quad ,
\end{equation}
\noindent respectively. Then, the difference between the maximum and the minimum is given by
\begin{equation} \label{A20a}
  \Delta F_2 / b f_{int} = \frac{(\frac{\pi}{2} - \theta_{min})}{sin(\theta_{min})} - cos(\theta_{min})   \quad ,
\end{equation}   
\noindent which provides the height of the barrier preventing that the system can move
from the angle $\theta_c$ to the angle $\pi - \theta_c$, which is needed for the droplet to
subsequently grow
increasing its baseline beyond $b$ at fixed angle $\pi - \theta_c$. Expanding equation (\ref{A20a})
in terms of the angle $\pi/2 - \theta_{min} = \alpha$, one gets 
$\Delta F_2 / b f_{int}  \simeq \frac{2}{3} (\alpha^{3})$.
For $\theta_{min} = 70^{\circ}$ ($\alpha = 20^{\circ}$) this leads to a barrier (in units of $k_B T$) of about $2.6$.

Actually, when $m_{coex} b H_b$ approaches $f_{int}$, then the angles where
the minimum and maximum of the free energy occur, merge at $\theta = 90^{\circ}$
(c.f. figure \ref{fig13} for $b H_b = 1.8$). However, already at a smaller field
(keeping $b$ constant) the barrier caused by the free energy maximum, $\Delta F_{2}$
given by equation (\ref{A20a}), will be small enough so that the second
nucleation event by which the angle grows from $\theta_c$ to $\pi - \theta_c$
can take place. Note that the analytical formula for the barrier can also be extracted
from equations (\ref{A17}), (\ref{A19}), and figure \ref{fig13} shows that long
before $\theta$ reaches $90^{\circ}$  it will be of order of a few $k_B T$ only. This argument  also
explains why the temperature dependence of the apparent angle (close to
$70^{\circ}$) where the depinning transition occurs is rather weak (see Table I):
the scale for the barrier is simply set by $b f_{int}$, and this quantity
does not vary strongly with $T$ for the choices we have made.
Thus when this barrier is small enough the instability that would occur
for $\theta = 90^{\circ}$ (where $\theta_{min}$ and $\theta_{max}$ merge) is preempted by the
jump of the angle $\theta$ from $\theta_{c}$ to $\pi - \theta_c$ .

\begin{table}
  \caption{List of the bulk fields ($H_b^{max}$, 3rd column) where the jump of the
    excess density indicating the formation of the liquid phase is observed,
    as measured for several choices of the temperature $t$ (1st column), given in units of the
    bulk critical temperature of the Ising model. Also, the interface tension 
    $f_{int}$ given by the Onsager exact solution \cite{46h} and the apparent angle
    $\theta_{min}$, as determined by using equation (\ref{A19}) are listed 
    in the second and fourth columns, respectively.  
    Data obtained by taking $b = 51$ for the length of the inhomogeneity.
    Note that these transition fields can be estimated only with a relative error
    of about one percent, and a similar error is expected for $\theta_{min}$.
    Both $f_{int}$ and $H_b^{max}$ are given in units of $J$.
  }
  \begin{tabularx}{\textwidth}{XXXX}
    \hline\hline\hline\hline
    $t$       & $f_{int}$       & $H_b^{max}$          & $\theta_{min}$ Eq.(\ref{A19})  \\
   0.30       & 1.92780       & 0.0354(2)         &  69.5$^{\circ}$  \\
   0.325      & 1.90190       & 0.0350(2)         &  69.8$^{\circ}$  \\
   0.350      & 1.87168       & 0.0339(2)         &  67.5$^{\circ}$  \\
   0.375      & 1.80873       & 0.0334(2)         &  70.4$^{\circ}$  \\
   0.3875     & 1.81848       & 0.0326(2)         &  66.1$^{\circ}$  \\
   0.4000     & 1.79873       & 0.0330(2)         &  70.0$^{\circ}$  \\   
   0.425      & 1.75625       & 0.0315(2)         &  66.2$^{\circ}$  \\
   0.430      & 1.74728       & 0.0308(2)         &  64.0$^{\circ}$  \\
   0.440      & 1.72851       & 0.0304(2)         &  63.8$^{\circ}$  \\
   0.445      & 1.71954       & 0.0302(2)         &  63.6$^{\circ}$  \\
    \hline\hline\hline\hline
  \end{tabularx}
\end{table}

On the other hand, one can change the height of the barrier by around
one order of magnitude just by taking a fixed temperature ($ t = 0.4$),
such as $f_{int} = 1.79873$, but varying the length of the inhomogeneity
$13 \leq b \leq 101$. In this way an increment of the
apparent angle of about $20^{\circ}$ is observed, as shown in Table II.
It is also obvious from equation (\ref{A19}) that solutions for $\theta_{min}$,
corresponding to pinned droplets, exist only for $m_{coex} H_b b /f_{int} < 1$ :
for larger fields droplet growth with time is not hindered by any barriers,
after they have been nucleated.

\begin{table}
  \caption{List of the the apparent angles $\theta_{min}$ (fourth column),
    as determined by using equation (\ref{A19}) for different choices
    of the length of the inhomogeneity $b$ (first column). Notice that
    the bulk fields ($H_b^{max}$, 3rd column) where the jump of the
    excess density indicating the formation of the liquid phase is observed,
    depend on $b$. Data taken at $t = 0.40$ so that the interface tension is given by 
    $f_{int} = 1.79873 $  according to the Onsager exact solution \cite{46h};
    however, the value of $b f_{int}$ (second column) that sets the height
    of the free energy barrier changes almost on order of magnitude
    for the choices of $b$ that are used.    
  }
  \begin{tabularx}{\textwidth}{XXXX}
    \hline\hline\hline\hline
    $b$       & $b f_{int}$       & $H_b^{max}$          & $\theta_{min}$ Eq.(\ref{A19})  \\
   13         & 23.38346        & 0.110(5)        & 52.7$^{\circ}$       \\
   17         & 30.57838        & 0.088(3)        & 55.8$^{\circ}$        \\
   21         & 37.77329        & 0.075(3)        & 61.1$^{\circ}$       \\
   25         & 44.96820        & 0.062(2)        & 59.5$^{\circ}$       \\
   31         & 55.76057        & 0.051(2)        & 61.5$^{\circ}$       \\
   39         & 70.15039        & 0.040(2)        & 60.1$^{\circ}$       \\
   45         & 80.94276        & 0.036(2)        & 64.2$^{\circ}$       \\
   51         & 91.73513        & 0.033(2)        & 70.0$^{\circ}$       \\
   75         & 134.90475       & 0.023(2)        & 73.5$^{\circ}$       \\
   101        & 181.67173       & 0.017(2)        & 72.7$^{\circ}$       \\
 \hline\hline\hline\hline
  \end{tabularx}
\end{table}

The smooth variation of $\theta$ from $\theta_c$ to $\pi - \theta_c$ with increasing droplet 
area as predicted by Lipowsky et al. \cite{66h,67h,68h} is a special consequence of the canonical 
ensemble, where the droplet volume (in $d = 3$) or the droplet area (in $d = 2$) is taken as a
fixed independent variable. In contrast,  
only part of the variation is realizable as a metastability effect in the grandcanonical 
ensemble, where $H_b$ is given. According to the theory outlined above, metastable
pinned droplets should exist only up to a "spinodal" where 
$\theta_{min} = \frac{\pi}{2}$. The spinodal field is then $H_b^{spin} = f_{int}/b m_{coex}$,
such that  $H_b^{spin} / H_b^{crit} = 1/sin(\theta_c)$
(see also equations  (\ref{eqhaa}) and (\ref{A19})).
However, it should be kept in mind that for systems with short-range
interactions "spinodals" are a somewhat ill defined concept \cite{2h}
and cannot be reached in practice.
In the context of nucleation phenomena, the present case
of a grandcanonical ensemble is the physically meaningful choice, of course.

For the understanding of the results observed in the simulations, it is hence
crucial to consider the combined effects of the primary nucleation event of the
wall-attached droplet and a further growth of this droplet.

We have made the hypothesis, that for this growth a "local equilibrium" assumption
holds, in particular near the point where the droplet-vapor  interface
meets the substrate.
This implies, for the case where the length $b_{drop}$ of the growing droplet still is
less than $b$, that we have $\theta = \theta_c$ for the contact line of the growing droplet
(see Figure \ref{FigA2} in Appendix A, top panel). However, for droplets that have $b_{drop} > b$, we have
$\theta = \pi - \theta_c$ (see Figure \ref{FigA2} in Appendix A, lower panel).
For fields $H_b < H_b^{spin} = f_{int}/(b m_{coex})$ metastable pinned droplets are predicted,
and a barrier $\Delta F_2$ for the ``depinning'' of these droplets could be estimated (Equation (\ref{A20a})).
The actual limit of stability of metastable pinned droplets, as seen in Figures \ref{fig7}, \ref{fig8}, is
somewhat smaller than $H_b^{spin}$: this happens because when the barrier $\Delta F_2$
is small, it can be overcome in a second nucleation event.

Of course, a perfect quantitative agreement of the predictions 
based on our phenomenological theory for pinned droplets with 
the corresponding simulation results should not be expected: (i) 
the mean-field like treatment of equations (\ref{eqang})-(\ref{A19})
disregards the huge statistical fluctuations that are 
present (Figure \ref{fig2}), (ii) the anisotropy of the interfacial
free energy should lead to some deviations of the shape of 
the pinned droplets from the circle cut, which should cause
some systematic deviations from the free energy plotted in 
figure \ref{fig13}. Also, the curvature of the interface 
may modify the effective surface tension.

\section{Nucleation Kinetics and Droplet Growth.}

Already in earlier work on studies of homogeneous nucleation
in bulk Ising models (see e.g.  \cite{31h} for a recent review) it has
been shown for conditions where the phase transformation is caused
by nucleation and growth of a single droplet one needs to distinguish
two very different time scales: The typical lifetime of the metastable
state, which is then simply inversely proportional to the nucleation rate; and
the time needed for the nucleated droplet to grow and essentially occupy the
total volume of the simulation box. However, often these processes are somewhat
confused by the crossover to the regime where during the phase transformation
many droplets are nucleated in different parts of the system and the lifetime of the
metastable state then is limited by this competitive growth of many droplets. This
latter regime is dominant when the simulation volume is relatively large
and $H_b$ is not so small, so nucleation becomes relatively easy \cite{2h}.

\begin{figure}[ht]
\includegraphics[width=10cm,clip]{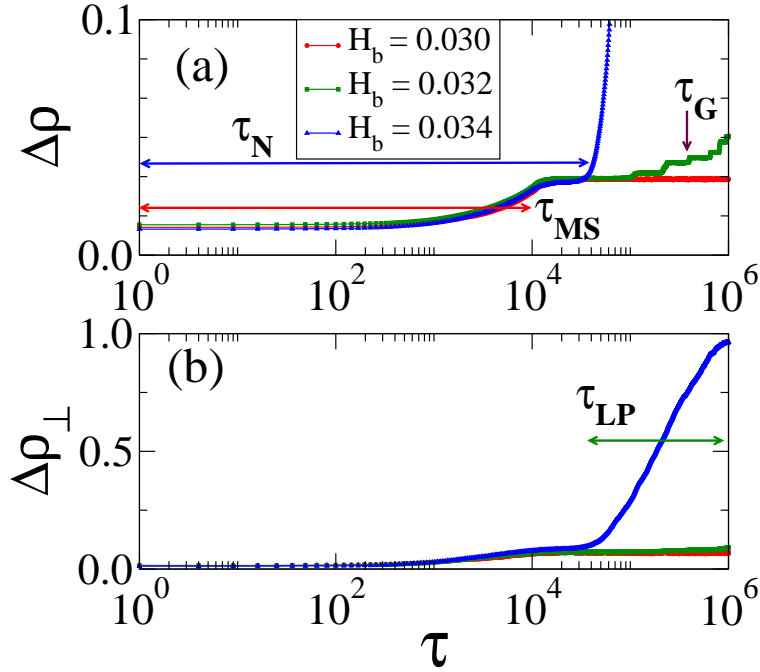}
\caption{\label{fig14}
a) Linear-log plot of the density {\it versus} time ($\tau$) as obtained for 
$t = 0.40$, $b = 51$, and different choices of the bulk field $H_b$, as
indicated. The horizontal lower and upper double arrows indicate the average times required to
achieve the metastable state and the nucleation time, $\tau_{MS} \simeq
10^4$ MCS, and $\tau_N \simeq 4\times 10^4$ MCS, respectively. Also, 
the vertical arrow shows a single growth event of an already nucleated droplet with 
a characteristic growth time given by $\tau_G \simeq 10^4$ MCS.  
Note that the fields $H_b$ included here are all slightly less than the
predicted ``spinodal'' (ultimate stability limit) $H_{b}^{spin} \simeq 0.0353$:
hence here droplet growth involves the second nucleation event, in which the
(small) barrier $\Delta F_{2}$ is overcome (see Equation (\ref{A20a}) and Figure \ref{fig13}).
b)  Linear-log plot of the linear density as measured at the center of the
sample and in the direction perpendicular to the wall ($\Delta \rho_{\bot}$)
{\it versus} time as obtained for the same parameters as in (a). 
The horizontal double arrow shows the average time needed to cover the 
whole sample for $H_b = 0.034$ with the liquid phase, $\tau_{LP} \simeq 8 \times 10^5$ MCS.    
Averages were taken over $386$, $453$ and $318$ different realizations for 
$H_b = 0.030$, $H_b = 0.032$, and $H_b = 0.034$, respectively.
More details in the text. 
}
\end{figure}

In the present work, conditions were chosen such that homogeneous nucleation
is not observable at all, and heterogeneous nucleation is restricted to the boundary
region of length $b$.  It then is rather straightforward to follow the growth of the single
droplet (Figure \ref{fig11}), and it turns out that the time intervals between
the snapshots of the growing droplet are indeed very small in comparison with the
nucleation times. In order to give further insight on the involved times
as well as on the growing and nucleation process of the droplets, Figure 
\ref{fig14} shows plots of the time evolution of both the total excess density $\Delta \rho$
due to the droplet (upper panel)
and the linear density excess  $\Delta \rho_{\bot}$ measured in the direction perpendicular to the wall
just at the center of the droplet. Each curve is the average over several hundred
individual time evolutions of the system.
In Figure \ref{fig14}(a) one can roughly
estimate the average time required by the system to achieve the metastable
state ($\tau_{MS}$), which for the case shown (i.e. $t = 0.40$ and $b = 51$) 
is $\tau_{MS} \simeq 10^{4}$ MCS. Choosing $H_b = 0.030$ no nucleation events
are detected during the observation time ($\tau_{obs} = 10^6$), and the curve
remains flat after achieving the metastable state. For $H_b = 0.032$ few
nucleation events are detected, and each of them shows up as an upward step
in the corresponding plot. The height of each individual step simply is the
inverse of the number of runs, since in each run when nucleation has occurred the
droplet grows fast (on the scale of $\tau_{N}$) to fill the available area.
Here, one can estimate the typical growth time ($\tau_G$)
required for each already nucleated droplet to expand over the whole sample,
namely $\tau_G \simeq 10^4$ MCS. However, for  $H_b = 0.034$ nucleation is
dominant and one can estimate $\tau_N \simeq 4\times 10^4$ (also by discounting
$\tau_{MS}$ one can get $\tau_N \simeq 3\times 10^4$ MCS). On the other hand,
the time evolution of the density per unit length as measured in the
direction perpendicular to the sample (Figure \ref{fig14}(b)), which shows
the development of the droplet in that direction, is fully
consistent with the above discussed scenario. Furthermore, here one can also
estimate the average time elapsed between the onset of nucleation and the 
achievement of a full liquid phase covering the whole sample, 
$\tau_{LP} \simeq 8\times 10^5$ MCS. Note that this averaged time results from 
the contribution of many growing events of already nucleated droplets, 
occurring at different times over a wide time interval, 
(actually 318 events for $H_b = 0.034$) each of them having a short lifetime 
of the order of $\tau_G \simeq 10^4$ MCS, as already discussed.

The individual nucleation events seen in figure \ref{fig14} actually all relate to
overcoming the barrier $\Delta F_2$ discussed in Figure \ref{fig13}, since the fields $H_b$
all are slightly below the stability limit $H_b^{spin}$.  

\begin{figure}[ht]
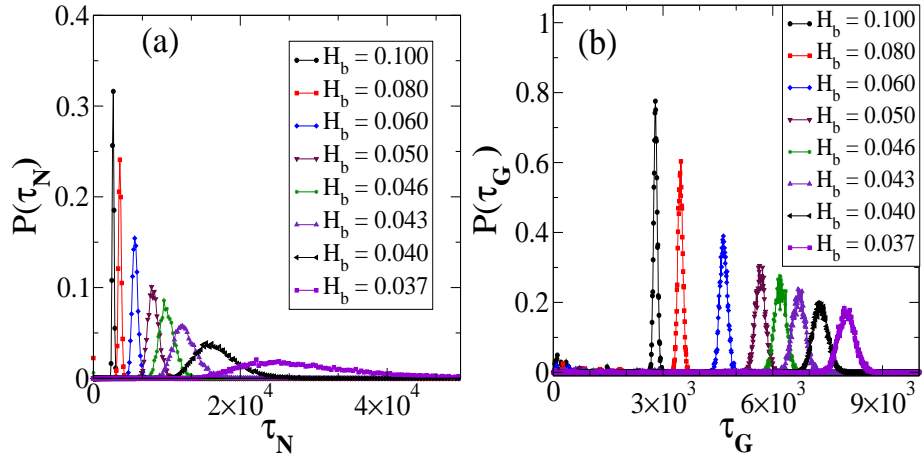

\includegraphics[width=6cm,height=6cm,angle= 0,clip]{Figure15a.eps}
\includegraphics[width=6cm,height=6cm,angle= 0,clip]{Figure15b.eps}
  \vskip 2.00cm
\caption{\label{fig15}  a) Plot of the nucleation time distribution function $P(\tau_N)$ 
{\it versus} time as obtained for different values of the bulk field $H_b$ as indicated.
Results correspond to $t = 0.40$, $b = 51$. Data averaged over $5 \times 10^{3}$ and
$15 \times 10^{3}$ different initial configurations for $H_b \geq 0.043$ and $H_b < 0.040$,
respectively. b) Plot of the growth time distribution function $P(\tau_G)$ 
{\it versus} time as obtained for the same choice of parameters as in a).  
}
\end{figure}

For a more quantitative analysis, we have also recorded both the nucleation time
distribution ($P(\tau_N)$) and the growth time distribution ($P(\tau_G)$) for
the case $t = 0.40$,  $b = 51$, and for different choices of $H_b$, as shown 
in  Figures \ref{fig15} (a) and (b), respectively.

\begin{figure}[ht]
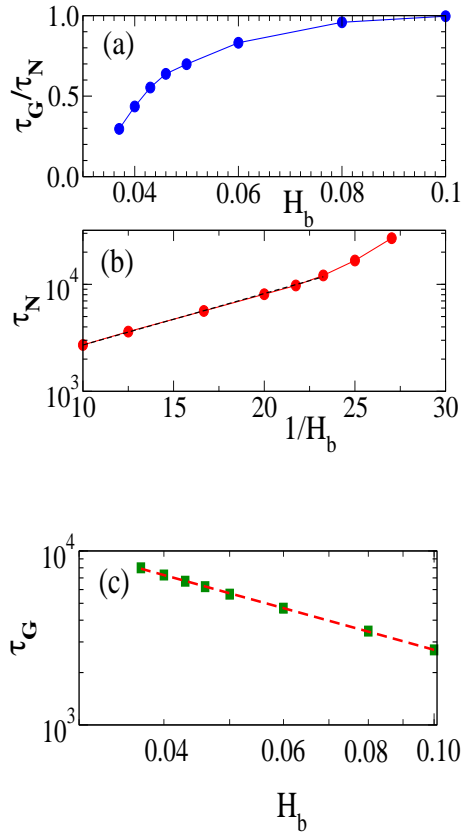

\includegraphics[width=6cm,height=6cm,angle= 0,clip]{Figure16ab.eps}
  \vskip 1.00cm
\includegraphics[width=6cm,height=4cm,angle= 0]{Figure16c.eps}
\caption{\label{fig16} a) Plots of the ratio 
$\tau_G / \tau_N$ {\it versus} $H_b$ for the case $b = 51$. The characteristic times are  obtained 
from estimations of  the peaks of the distributions shown in 
Figure \ref{fig15}.
b) Log-linear plot of $\tau_N$  {\it versus} $1/H_b$. The fit of the data within the linear regime
($H_b > 0.04$) shown by a dashed line yields 
$\Delta F^{*} H_b /k_B T = 0.111 $.  
c) Plot of $\tau_G $ {\it versus} $H_b$ on logarithmic scales. The straight
line corresponds to the best fit of the data with slope $-1.08(8)$ in agreement with
the expected theoretical dependence, namely $\tau_G \propto H_b^{-1}$.
More details in the text.
}
\end{figure}

These choices all refer to $H_b > H_b^{spin}$, and hence for them the barrier
$\Delta F_2$ does no longer occur.
A simple comparison of
both figures indicates that the characteristic times, as estimated from the
location of the peaks of the distributions, are roughly of the same order
for larger fields ($H_b \ge 0.08$), while $\tau_N > \tau_G$ in the opposite
limit. In fact, figure \ref{fig16} (a) shows the monotonic 
increase of the ratio $\tau_G / \tau_N$ when it is plotted {\it versus} $H_b$,
spanning the range $0.25 \leq  \tau_G / \tau_N \leq 1$.
Also, Figure \ref{fig16} (b) shows a plot of $\tau_N$ {\it versus}
$1/H_b$. Nucleation theory predicts  
\begin{equation} \label{eqtau}
ln(\tau_N) \propto \frac{\Delta F^{*}}{k_B T} = \frac{\pi}{2} f_{int}^{2}
\frac{1} {\Big(\rho^{\rm coex}_\ell - \rho^{\rm coex}_v \Big)  H_{b}}
\frac{f_{VT}}{k_B T}
  \quad .
\end{equation}
However, the curvature of the log-linear plot indicates that only part of the
chosen region of fields is in the regime where the barrier due to the
heterogeneous nucleation on the inhomogeneity controls the kinetic exclusively;
in fact, when $H_b$ approaches $H_b^{spin}$ a slowing down related to the
barrier $\Delta F_2$ that occurs for $H_b < H_b^{spin}$ may be present. 
The best fit of the data of Figure \ref{fig16} (b), within the linear regime,
yields $\frac{\Delta F^{*} H_b }{k_B T} = 0.111 $.
This number is smaller than the theoretical expectation given by
$ \frac{\pi}{2} f_{int}^{2} \frac{1} {\Big(\rho^{\rm coex}_\ell - \rho^{\rm coex}_v \Big)} \frac{f_{VT}}{k_B T}$
(see equation (\ref{eqtau})),
obtained by taking $f_{int} = 1.7987$ \cite{46h}, which yields
$\frac{\Delta F^{*} H_b }{k_B T} = 0.231$ for $\theta_c = 34^{\circ}$
in the Volmer-Turnbull factor (Equation (\ref{A15})).
It is a subtle issue to understand where this discrepancy 
of about a factor two in the effective barrier height comes from.
On inmediate thought concerns the curvature dependence of the 
interfacial free energy $f_{int}(R)$. In $d = 2$ there is 
evidence from field theoretical calculations \cite{x1}, Monte
Carlo simulations of cluster-size distributions \cite{x2},
and analysis of the two-phase coexistence \cite{x3} that
\begin{equation}
\frac{f_{int}(R)}{f_{int}(\infty)} = 1 + \frac{5}{4\pi f_{int}(\infty)} \frac{\ln(R)}{R} +
        \frac{const}{R} \quad ,
\end{equation}
\noindent where the constant in the last term on the right-hand side
is non-universal, while the prefactor $ \frac{5}{4\pi}$ of the logarithmic 
term is universal. If only this correction would be taken into account,
the interfacial tension be enhanced by a factor $1 + 0.221  \frac{\ln(R)}{R}$,
which for typical values of $R$ (e.g. $R = 16$) is an enhancement of about $4\%$.
Neither the magnitude nor the sign of this effect can account for the observed 
discrepancy. Actually a more plausible assumption is that our estimate  of the 
contact angle $\theta_c$ and hence the factor $f_{VT}(\theta_c)$ is an overestimate.
Since  $f_{VT}(\theta_c) \approx \frac{4}{3 \pi} \theta^3$ a decrease of $\theta_c$
by a few degrees already suffies to reduce  $f_{VT}(\theta_c)$ by a factor of two,
e.g.  $f_{VT}(27^{\circ})/f_{VT}(34^{\circ}) \simeq 0.51$.
Moreover the equation for $f_{VT}(\theta_c)$ holds only for circle-cut shaped 
droplets, and the effect of anisotropy cuasing somewhat non-circular
shapes (see figure \ref{FigA1} in the Appendix A) of the droplet
on $f_{VT}(\theta_c)$ still needs to be clarified. Also, the curvature
of the plot shown in figure \ref{fig16} b) may be taken as an indication
that it is questionable whether the asymptotic region where the theory 
holds has been reached.  Thus clearly the conclusion emerges that 
in spite of the simplicity of the Ising model still more work is needed 
to understand there heterogeneous nucleation quantitatively.  
  
\begin{figure}[ht]
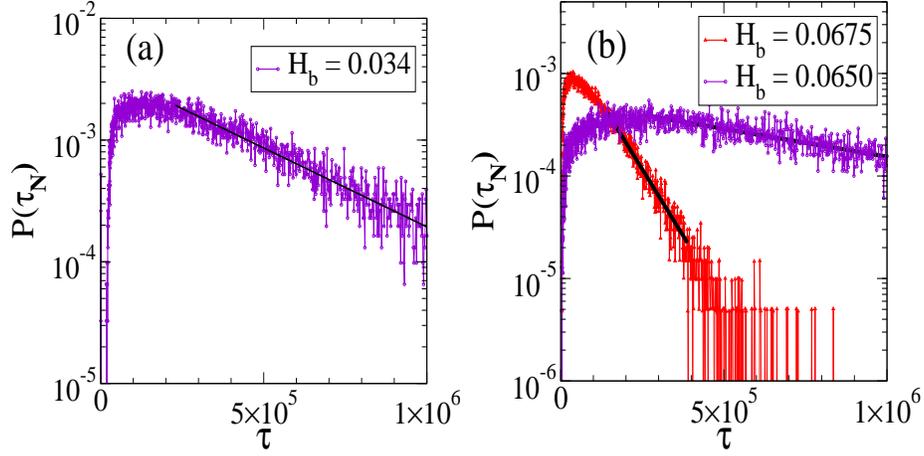

\includegraphics[width=6cm,height=6cm,angle= 0]{Figure17a.eps}
\includegraphics[width=6cm,height=6cm,angle= 0]{Figure17b.eps}
\caption{\label{fig17} 
Log-linear plots of  $P(\tau_N)$ {\it versus} time as obtained for 
$t = 0.40$, and different choices of the bulk field as indicated.
a) Results corresponding to $b = 51$. After a waiting 
time ($\tau_w \simeq 2\times 10^4$ MCS in this case) the distribution exhibits
an exponential decay with a characteristic nucleation time $\tau_N \simeq 3.36
\times 10^5$  MCS, as obtained 
from the slope of the fitted curve 
as shown by means of a full line.   
b) Results corresponding to $b = 25$. By following the procedure described
in a), the fitted characteristic nucleation 
times are $\tau_N \simeq 7.78 \times 10^5$  MCS, and  
$\tau_N \simeq 8.40 \times 10^4$ MCS, for $H_b = 0.0650$ 
and $H_b = 0.0675$, respectively. In both cases the slopes of the fitted
curves are shown by means of full lines.
Note that in both cases the fields are slightly less than $H_b^{spin} = 0.072$,
and hence the combined effect of heterogeneous nucleation on the wall
inhomogeneity and depinning of the metastable droplet matters.
}
\end{figure}

Figure \ref{fig16} c) shows a log-log plot of $\tau_G$ {\it versus}
$H_b$ to show that the growth time $\tau_G$ scales inversely with $H_b$, as expected.
On the other hand, figure \ref{fig17}
shows log-linear plots of  $P(\tau_N)$ {\it versus } $\tau_N$, for different
choices of the length of the inhomogeneity $b$ and the bulk field $H_b$,  
demonstrating an exponential distribution for the long times, as theoretically
expected \cite{35h,54h}.

Of course, it would be interesting to explore the kinetics of heterogeneous nucleation
systematically for a wide range of $b$, but due to excessive needs for computer
time this has not been attempted. So, we have only studied one other choice, $b = 25$.
For $b = 25$ one has $H_b^{spin}/J \simeq 0.060$, so the
measurements performed slightly above $H_b^{spin}$ indicate that $\tau_N$ 
depends strongly on $H_b$, and decrease for larger values of $H_b$, as
expected. The fact that $\tau_N$ for $b= 51$ and $H_b/J = 0.034$ is smaller than 
for the case $b = 25$ and  $H_b/J = 0.064$ means that one has smaller
nucleation times for larger inhomogeneities, since the larger the inhomogeneity
is, the less tendency is found that the nucleated droplet gets pinned.

\section{Conclusions.}

In this work, we have considered the effect of a chemical inhomogeneity on heterogeneous
nucleation on a flat substrate. As a generic case, we have considered the simple Ising
lattice gas system in $d=2$ dimensions, where the flat substrate is just a straight line,
and the chemical inhomogeneity is represented by a positive boundary field $H_{w3}$
(favoring the liquid phase of the lattice gas) acting on a part of this line of length $b$,
while in the remaining part of this lower boundary of the system a boundary field
$H_{w2} = - H_{w3}$  is applied  favoring the vapor phase of the lattice gas.
For the Monte Carlo simulations of our model, we choose a boundary with finite
length $M$ and periodic boundary condition in the direction parallel to this boundary,
while in the direction perpendicular to this boundary a finite linear dimension $L$ is used,
and at the upper boundary a negative field $H_{w1} = H_{w2}/4$ acts, to stabilize the vapor
as a bulk phase of the system in the absence of a bulk field, $H_b = 0$.
For conditions of partial wetting, the density inhomogeneity in the lattice gas caused
by the chemical inhomogeneity then extends only over a distance of the order of one lattice
spacing, for $H_b = 0$ (Fig.\ref{fig3} a)). However, when  a small field $H_b > 0$ is applied,
the vapor phase chosen as the initial state of the system becomes metastable, and the
structure of the density inhomogeneity caused by the chemical inhomogeneity, Figs. \ref{fig3}-\ref{fig9},
as well as the decay rate of the metastable state due to nucleation of a boundary-attached
droplet and its growth Figs. \ref{fig10}-\ref{fig16}, are the subjects of investigation.
Conditions are chosen such that neither homogeneous nucleation in the bulk nor nucleation
starting in the boundary regions favoring the vapor phase can ever be observed.

We perform for each choice of temperature, width $b$ of the chemical inhomogeneity, and fields $H_b$,
many hundred equivalent Monte Carlo runs, differing by the pseudorandom numbers used to realize
the time evolution of the Monte Carlo sampling process. By using over $20 \times 10^6$ Monte
Carlo steps (MCS) per site, we are able to reliably  estimate various relaxations times and
their distributions (Figs. \ref{fig15}-\ref{fig17}) over $6$ decades of time. The initial stages of the
relaxation process are characterized by the equilibration of the metastable state,
after the field $H_b$ has been switched on at time $\tau = 0$, taking a time $\tau_{MS}$.
If $H_b$ is small enough, e.g. $H_b  \leq 0.030$ for the choice $t=0.40$, $b=51$, no decay of the
metastable state is observed, which implies that the nucleation time $\tau_N$ exceeds the observation
time $\tau_{obs}$.

If we would study heterogeneous nucleation on a chemically homogeneous boundary
of linear dimension $M$, the nucleation time (for the regime of fields where single-droplet
nucleation matters) would be related to the nucleation rate $J^{het}$ by $\tau_N = (M J^{het})^{-1})$.
When a nucleation event has occurred, it takes a time $\tau_G$ for the critical droplet to grow
until the whole (finite) system has transformed; only when $\tau_N  \gg \tau_G$ is the
phase transformation triggered by single nuclei the dominant process. In the regime where
$\tau_N$ and $\tau_G$ are comparable the simultaneous growth of multiple nucleated droplets
needs to be considered, making separate estimations of $\tau_N$ and $\tau_G$ difficult.
By choosing our geometry with a chemical inhomogeneity, we extend the regime where the
transformation triggered by single nuclei is the dominant process: note that the baseline of
the critical droplet is $b_{drop}^{\ast} = 2 R^{\ast} \sin(\theta_c)$, where $R^{\ast}$ is the critical droplet
radius and $\theta_c$ the contact angle, assuming droplets of circle-cut shape;
only when $b_{drop}  \ll b$, phase transformations affected by nucleation of multiple droplets and
their competitive growth could matter. The detailed analysis of our observed phase transformation
events has allowed us a separate analysis of the distributions of $\tau_N$ and $\tau_G$;
as theoretically expected, the growth rate of supercritical droplets is proportional to $H_b$,
and hence $\tau_G \propto 1/H_b$ (Fig. \ref{fig16}c) ), while $\tau_N$ varies exponentially
with $1/H_b$, $\ln \tau_N  \propto 1/H_b$ (Fig.\ref{fig16}b)), as expected from nucleation
theory in $d=2$ dimensions. Unfortunately, only a very small range of $H_b$, much less than
a decade (Fig.\ref{fig16} b)), is available when the time scales for nucleation and growth
are well separated.

Very interesting behavior was found for the metastable regime, where during observation
time $\tau_{obs}$ no phase transformation occurs. On general grounds one can predict that
then the chemical inhomogeneity causes an excess density $ \Delta \rho$ in the system,
which exhibits a scaling behavior $ \Delta \rho = b^2 \tilde{M}(t, bH_b)$, Eq.(\ref{eqhqyq}),
the effective droplet height $h_{eff}/b$ similarly is a function of the product $b H_b$
only (Fig. \ref{fig3} c)), as well as the effective contact angle $\theta_{eff}$ (Fig. \ref{fig7} b)).
We hence identified two regimes: for very small values of $H_b$ such that $b_{drop} > b$,
nucleation of droplets with the contact angle $\theta_c$  “ preferred “ by  the chemical
inhomogeneity still is geometrically impossible, it does not matter how large observation
times are chosen. In fact, in this regime critical droplets would have the shape as
shown in Fig. \ref{FigA2} (lower part), their baseline $b_{drop}$ extending beyond $b$ and
the contact angle being $\pi - \theta_c$, but the corresponding nucleation barriers correspond
to astronomically large nucleation times, and hence are of no interest here.
In this regime, thermal fluctuations allow only the occasional formation of subcritical nuclei
with $R < R^{\ast}$, of circle cut shape with contact angle $\theta_c$.  Thus the average effect
of such fluctuations is measured by the scaling function $\tilde{M}(t, b H_b)$ for $H_b < H_b^{crit}$,
where $H_b^{crit}$ can be estimated as $H_b^{crit} = \sin(\theta_c) f_{int}/(b m_{coex})$,
when we ignore anisotropy effects on the interfacial tension in the lattice gas model.
In any cases $H_b^{crit}$ is the smallest field where a droplet with the correct contact angle
$\theta_c$ fits to the chemical inhomogeneity. For $H_b^{crit} < H_b < H_b^{spin}$ we may
encounter pinned droplets, having a baseline of length $b$, with contact angles exceeding
the equilibrium value, $\theta_c < \theta < \pi/2$ (cf. Fig. \ref{FigA2}). Assuming that
these droplets still have circle cut shape,  we have predicted that these droplets become
unstable for $H_b^{spin} = f_{int}/(b m_{coex})$, and we have obtained an approximation
for $\tilde{M}(t, bH_b)$ in this regime (Figure \ref{fig9} a)).  Note that further growth of
the droplets with baseline $b_{drop} > b$ requires that the contact angle grows up to
$\pi - \theta_c$, and in the regime $H_b^{crit} < H_b < H_b^{spin}$ this is hindered by
a free energy barrier $\Delta F_2$, see Figure \ref{fig13} and Equation (\ref{A20a}).
Since this barrier is only of the order of a few $k_BT$ when $\theta$  has reached about
$70^{\circ}$, pinned droplets with $70^{\circ} < \theta < \pi/2$ actually were not observed;
droplets with a shape as sketched in the lower part of Figure \ref{FigA2} then appear in a second nucleation
event, and grow to complete the phase transformation. So the actual limit of metastability
$H_b^{max}$ of pinned droplets is somewhat smaller than $H_b^{spin}$, e.g. (for $b = 51$)
$H_b^{max} b \simeq 1.71$ while $H_b^{spin} b \simeq 1.7987$.

\clearpage
\begin{figure}[ht]
  \includegraphics[width=6cm,height=5cm,clip]{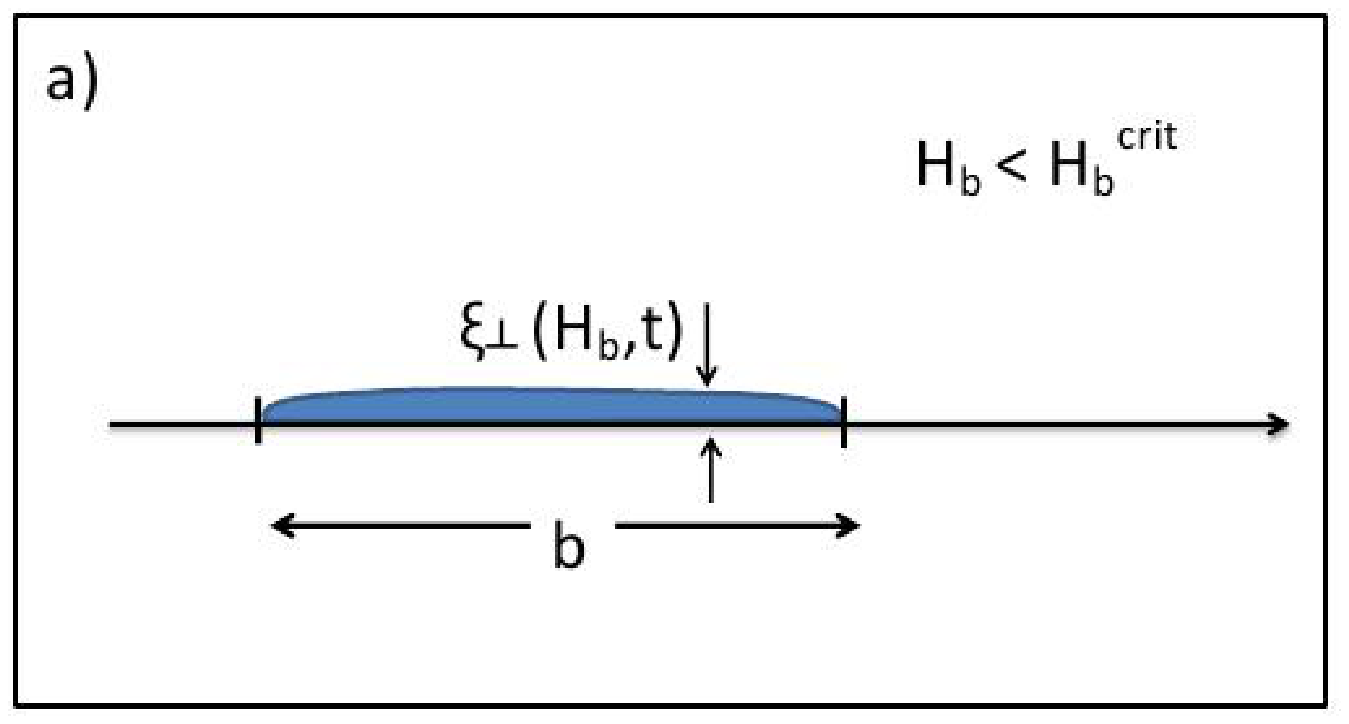}
  \includegraphics[width=7cm,height=8cm,clip]{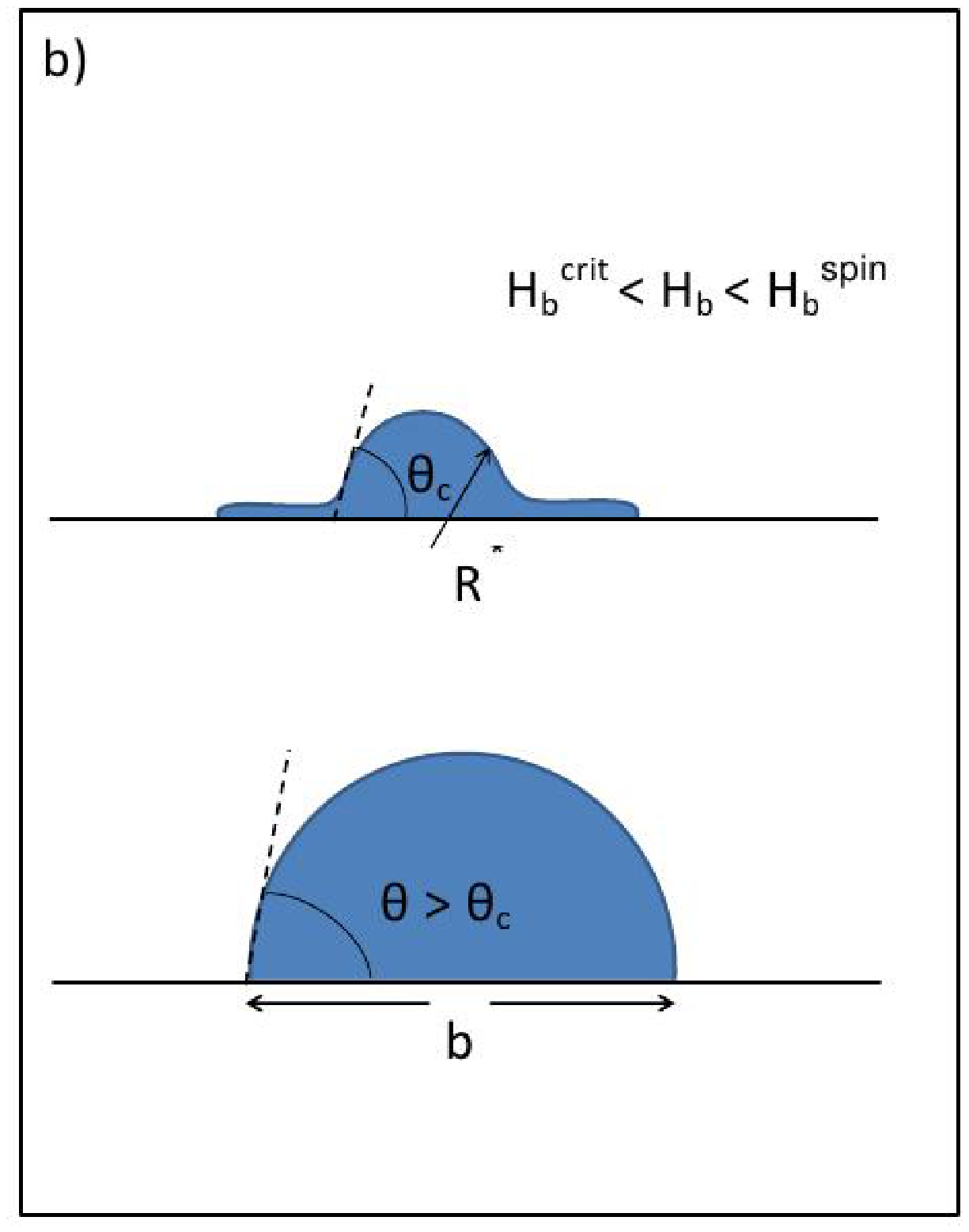}
  \includegraphics[width=5cm,height=7cm,clip]{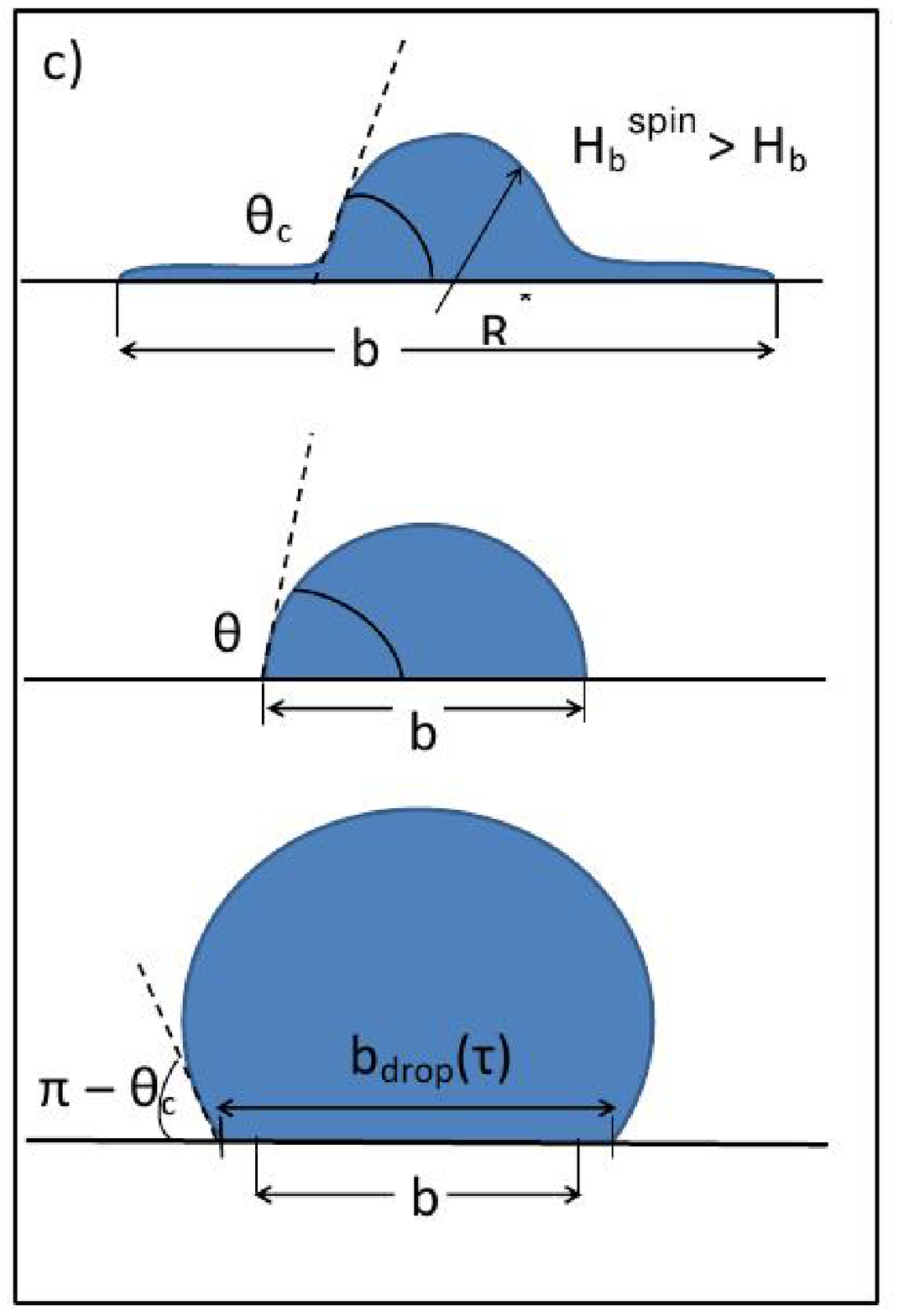}
  \includegraphics[width=5cm,height=5cm,clip]{Figure18d.eps}
  \caption{\label{fignew} See figure caption below.}
\end{figure}
\clearpage
Figure \ref{fignew} Caption. Sketches summarizing qualitatively the different 
regions of wall-attached droplets for the case of an inhomogeneity of linear
extension $b$ identified in this work.
Case a) shows the range where the bulk field $H_b$ is very small,
namely $H_b < H_b^{crit}$, ( $H_b^{crit} = f_{int}sin(\theta_c)/(b m_{coex}$)).
Then the critical droplet having a contact angle $\theta_c$ would require 
a baseline larger than $b$, since then $R^*$ exceeds $b/(2 sin(\theta_c))$,
see the case $H_b = 0.030$ in part d). No nucleation then is possible,
and due to the average effect of subcritical droplets a density excess 
$\Delta \rho \propto b \xi_{\bot}$ occurs on the inhomogeneity.
Case b) shows the regime $H_b^{crit} < H_b < H_b^{spin}$ ($H_b^{spin} = f_{int}/(b m_{coex}$)),
where critical droplets of radius $R^*$ and contact angle $\theta_c$ with baseline
$b_{drop}^* < b$ are nucleated, and their radius grows until their baseline is equal to $b$.
Then these droplets can lower their free energy further by increasing  their
contact angle from  $\theta_c$ to  $\theta_{min}$. These metastable 
pinned droplets are characterized by $sin(\theta_{min}) = m_{coex} H_b b/f_{int}$.
The case $H_b = 0.075$ in part d) illustrate the corresponding $\Delta F(R)$ in the regime
where $\theta = \theta_c$.
Case c) shows the behavior for $H_b^{spin} < H_b$, e.g. the case  $H_b = 0.20$ in part d),
where the critical droplet nucleates with contact angle $\theta_c$ has such a
small radius  $R^*$ and corresponding baseline, that after growth to the baseline $b$     
the increase of the contact angle is no longer pinned, and when the contact angle 
$\pi - \theta_c$ has been reached, the droplet can grow further with this contact angle 
and increasing thereby its contact line $b_{drop}(\tau)$ beyond $b$ with increasing time.
Thus, the two critical fields  $H_b^{crit}$ and $H_b^{spin}$ simply correspond to the cases
$R^* = b/(2 sin(\theta_c))$ and $R^* = b/2$, respectively. Note that in panel c)
the length of the inhomogeneity in the upper sketch was taken a factor two larger 
than in the medium and lower sketches for the sake of clarity. 
Panel d) shows plots of the excess free energy relative to the wall without droplet
$\Delta F(R)$ {\it vs} $R$ (see equation (\ref{A12}), as obtained for different fields
corresponding to the regimes shown in panels a)-c). Theoretical curves are obtained by taken
$\theta_c = 35^{\circ}$ and $f_{int} = 1.7987$ in order to illustrate the expected 
behavior for $t = 0.40$. The chosen fields are then suitable to describe the
case of an inhomogeneity of extension $b = 21$, such that  $b/(2 sin(\theta_c)) = 18.31$,
see the vertical dashed line. The values of $R^*$ corresponding to the selected
fields are shown along the horizontal axis.  
More details in the text.


Thus, an unexpectedly rich behavior concerning nucleation at a chemically
  inhomogeneous substrate has been found within the context of the grandcanonical ensemble
  used in our calculations (i.e. when the pressure of the fluid or equivalently the magnetic field
  of the Ising model is given as a control parameter). We have shown that this scenario differs from that
  corresponding to the canonical ensemble (i.e. when the volume in $d = 3$ or the area
  in $d = 2$ of the droplets is taken as a control parameter). In order to acquaint the reader
  with a clear description of the relevant findings reported in this paper, we have summarized
  and discussed our results in figure \ref{fignew}. In this way we addressed the relevant regimes
  encuntered in our study performed in the framework of the grandcanonical ensemble:
  a) The regime $H_b < H_b^{crit}$ where no nucleation is possible. b) The regime 
  $H_b^{crit} < H_b < H_b^{spin}$, where the droplet grows with contact angle $\theta_c$ until
  its baseline matches the length of the inhomogeneity, and then subsequently grows by
  keeping its baseline constant but increasing the contact angle. Finally, the regime c)
  corresponds to larger fields $H_b > H_b^{spin}$ that lie beyond the stability limit,
  so that the droplets can grow with baseline larger than the length of the inhomogeneity
  and contact angle $\pi - \theta_c$. Furthermore, all these three regimes are
  properly identified with the corresponding free energy functions $F(R)$ shown in panel
  d) of figure \ref{fignew}.

\clearpage

\begin{figure}[ht]
 \includegraphics[width=14cm,clip]{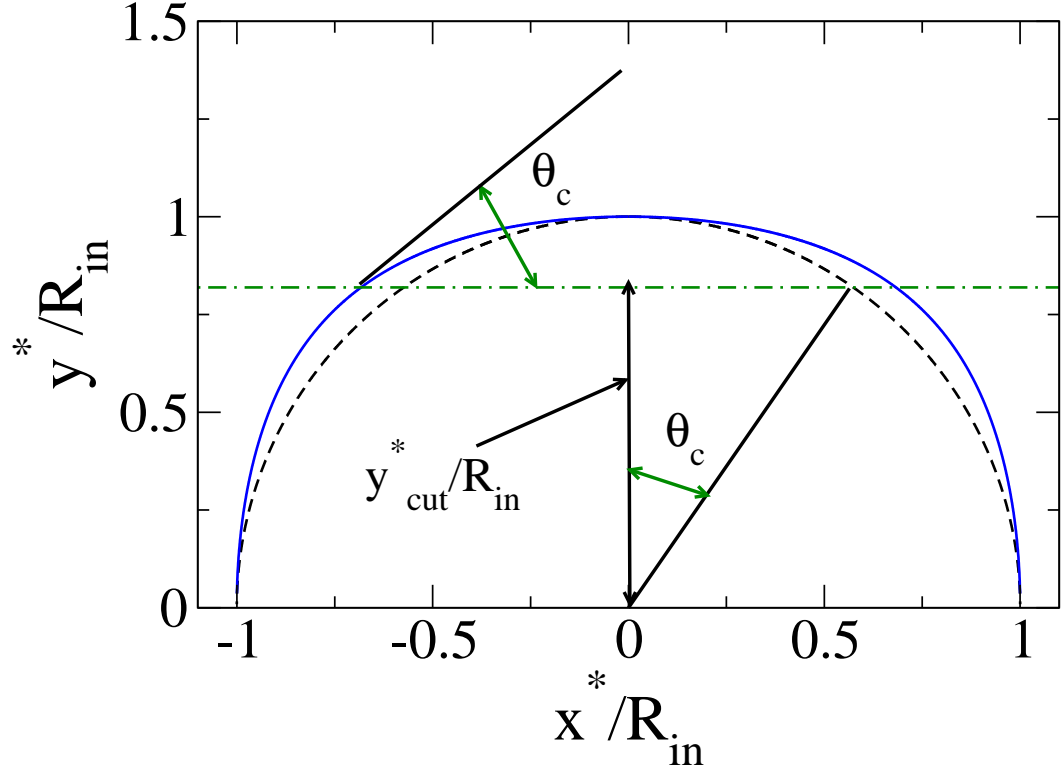}  
\caption{\label{FigA1}
The full line shows the shape of the droplet given by
the equation (\ref{A6}), while the dashed line shows the inscribed
circle of radius $R_{in}$ given by equation (\ref{A7}) and
evaluated at $t = T/T_{cb} = 0.40$, i.e.  $R_{in} = 1.98094$. 
The horizontal dashed-dotted line shows the location of the
circle cut line placed at a distance $y^{*}_{cut}/R_{in} = cos(\theta_c)$
from the origin. The contact angle of the droplet at the intersection between
the circle and the cut line, which is the angle that the droplet 
makes with the substrate, is also shown (notice that $\theta_c = 35^{\circ}$
has been the choice in this example). 
}
\end{figure}

\begin{figure}[ht]
  \includegraphics[width=7cm,height=4cm,clip]{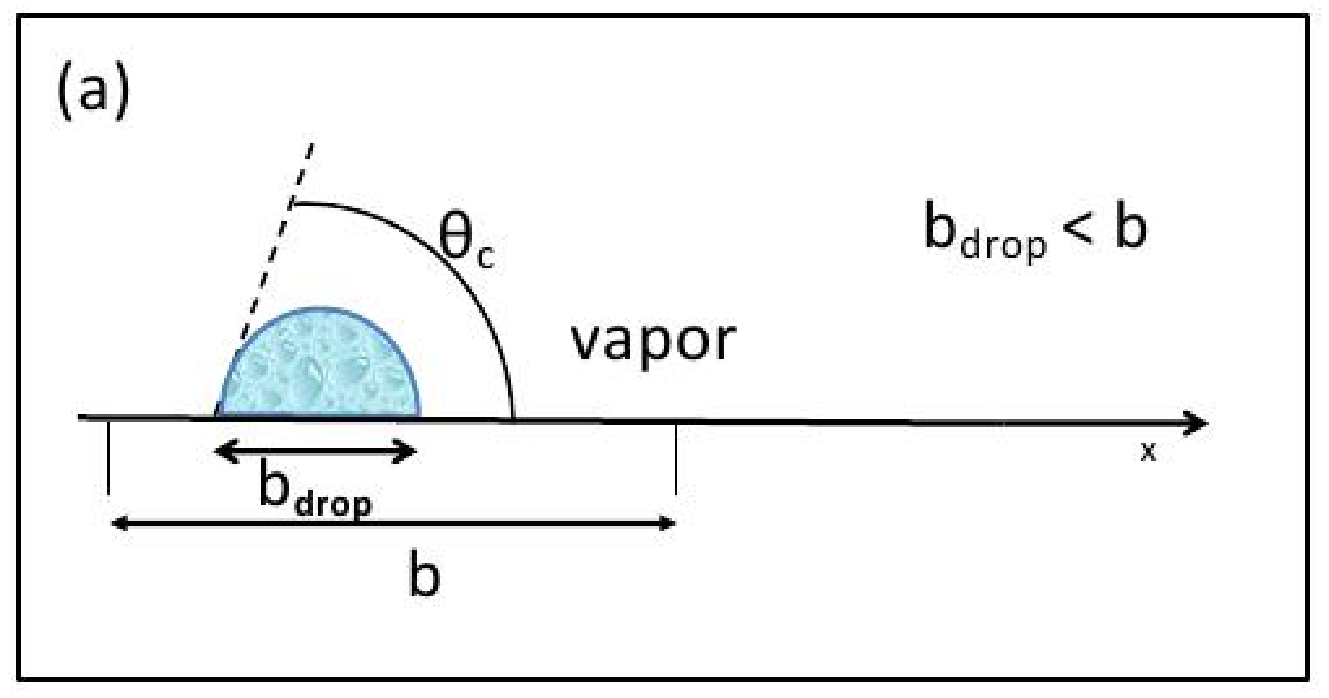}
  \includegraphics[width=7cm,height=4cm,clip]{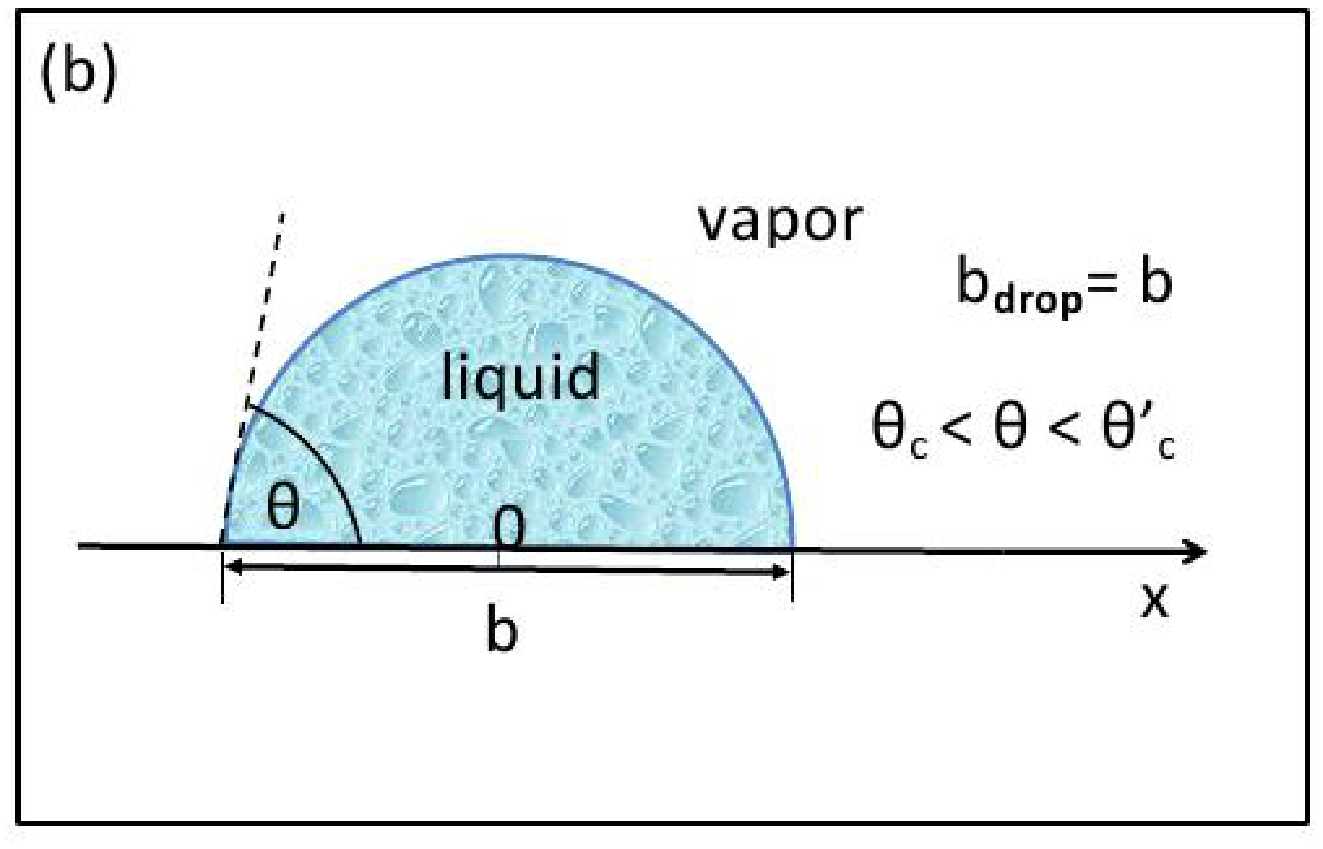}
  \includegraphics[width=7cm,height=4cm,clip]{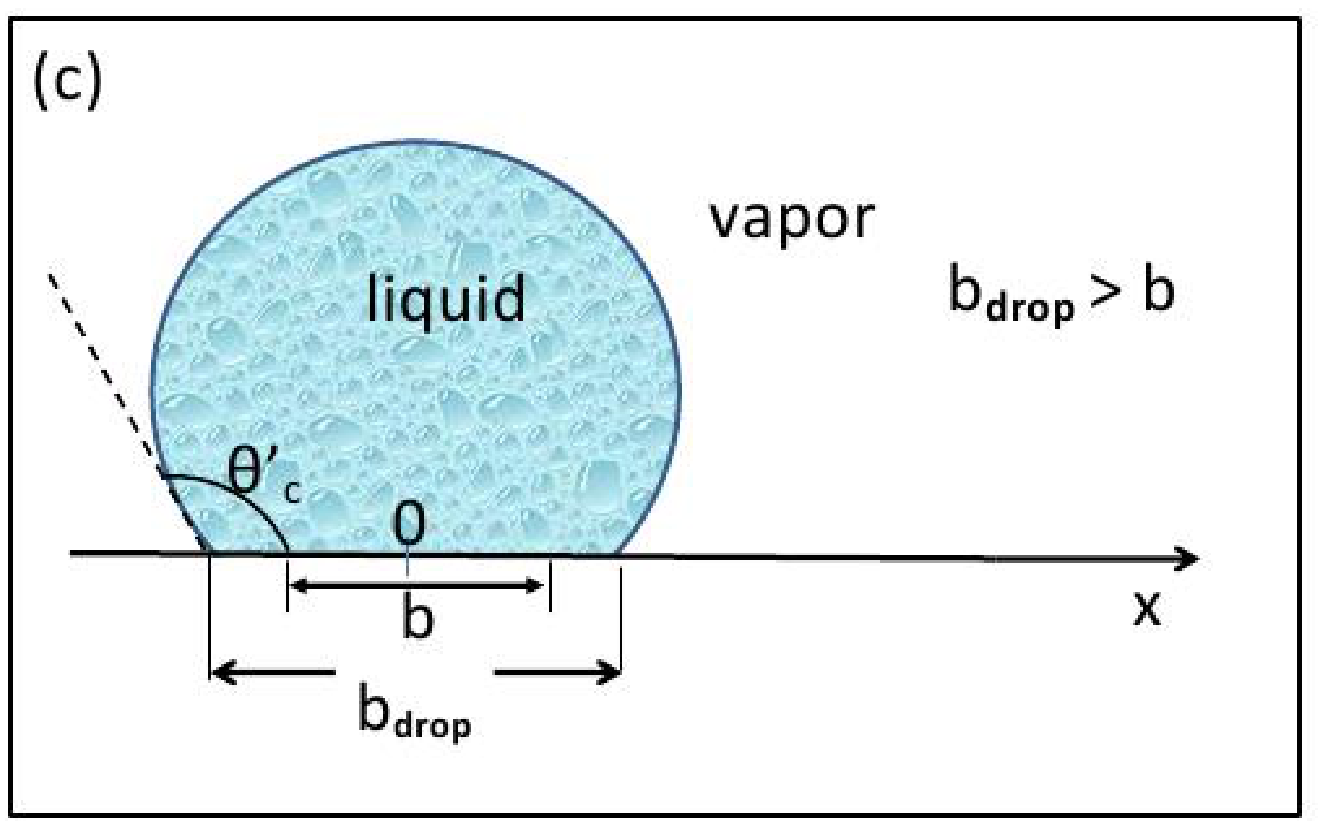}
  \caption{\label{FigA2} Sketches of droplets corresponding to
    the three relevant regimes considered by Lipowsky et al. \cite{66h,67h,68h},
    conveniently adapted to our two-dimensional case, namely: 
    I) $b_{drop} < b$ (upper panel); II) $b_{drop} = b$ (medium panel); and III) $b_{drop} > b$ (lower panel),
    where $b_{drop}$ is the
    baseline of the droplet in contact with the substrate, $b$ is the length of
    the heterogeneity, and
    $\theta_c$ is the contact angle. More details in the text.
}
\end{figure}

\begin{figure}[ht]
\includegraphics[width=14cm,clip]{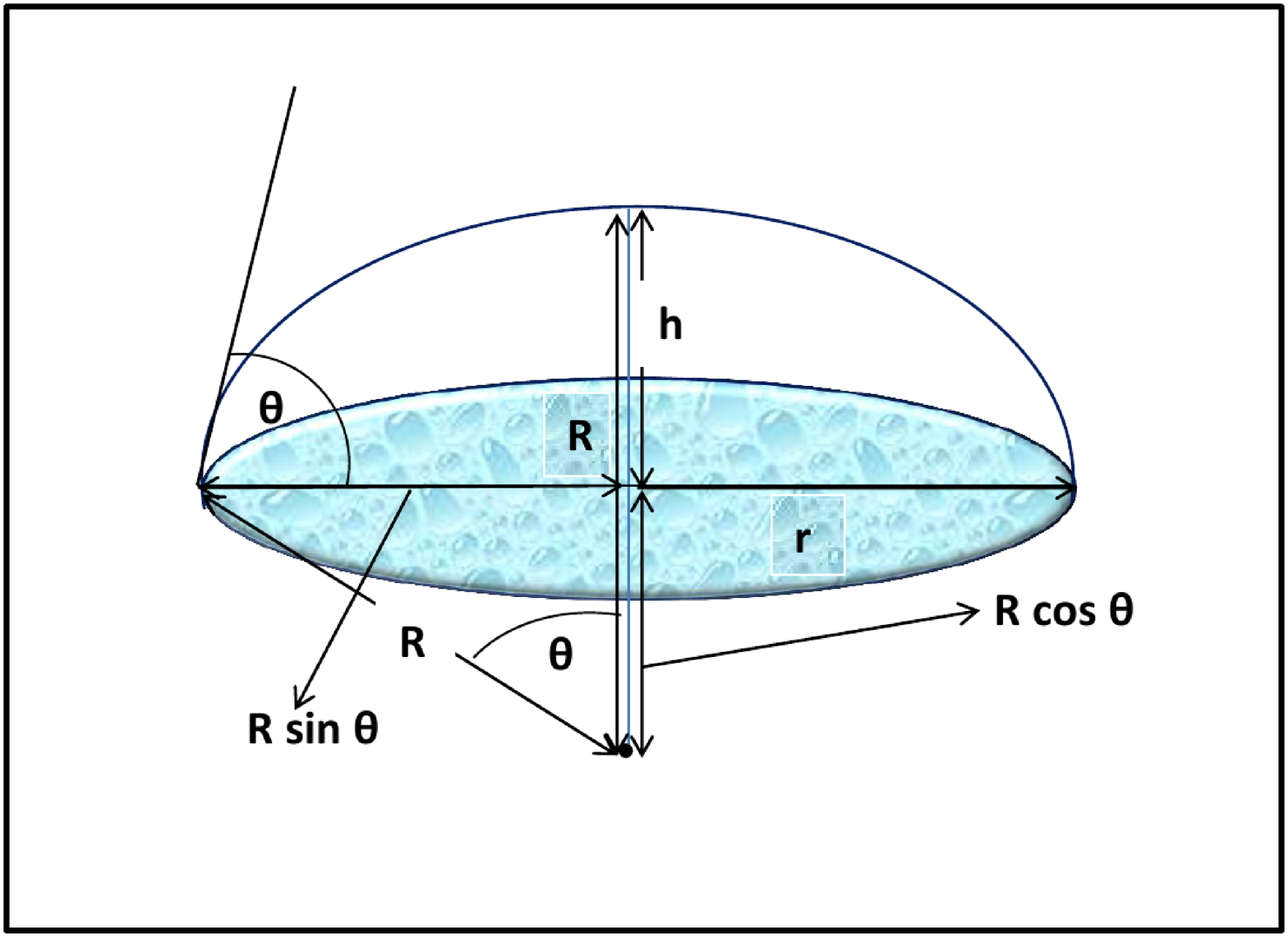}
  \caption{\label{figB1} Geometrical description used to construct a
spherical cap droplet, where $\theta$ is the angle that the droplet 
makes with the substrate. More details in the text.
}
\end{figure}

\begin{figure}[ht]
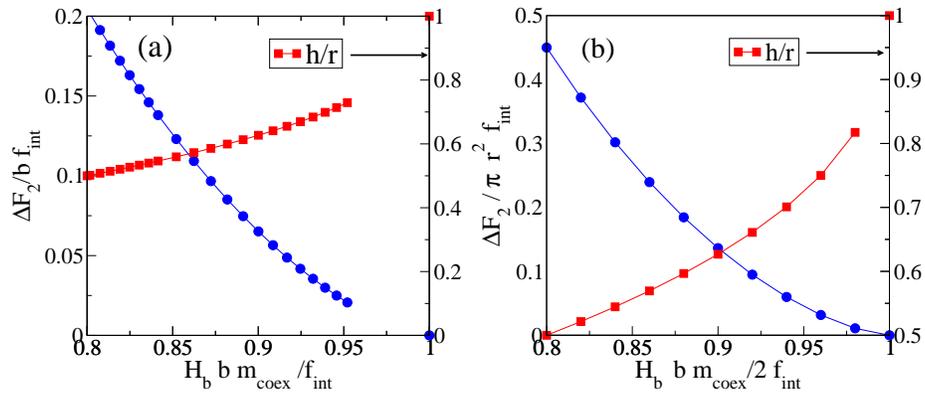

  \includegraphics[width=6cm,clip]{Figure22a.eps}
  \includegraphics[width=6cm,clip]{Figure22b.eps}
\caption{\label{figB2}
Comparison of the free energy barrier of the
depinning transition of pinned droplets for the two- and the three-dimensional cases
shown in panels a) and b), respectively. For $d = 3$ we assume $b = 2 r$, see
figure \ref{figB1}. The left ordinate scales refer to the scaled free energy barrier
and the right scales refer to $h/r$, respectively. More details in the text. 
}
\end{figure}

\appendix
\section {Heterogeneous nucleation on homogeneous and chemically inhomogeneous
substrates: Phenomenological Theory.}

We consider here a fluid in $d = 2$ dimensions exposed to a 
(one-dimensional) flat wall, under conditions of partial wetting.
For simplicity, we shall consider explicitly only the nearest-neighbor 
lattice gas (Ising) model and the wall is then oriented along the $x-$direction 
of the lattice; the Ising spins adjacent to this wall then have a row of 
missing neighbors but experience a surface field $H_{w3}$ chosen such that the liquid 
phase of the lattice gas model is favored; unlike Figure \ref{fig1} we consider
first the case where this boundary field is homogeneous independent of the coordinate $x$.
In the semi-infinite system at zero bulk field $H_b = 0$, the vapor phase can be stable
(due to a suitable boundary field at $y = L \rightarrow \infty$), and near the boundary
where $H_{w3}$ acts only a liquid film of the average thickness of order $\xi_{\bot}$
is stabilized; $\xi_{\bot}$is the perpendicular correlation length of the wetting 
transition, and in the regime of partial wetting this is a microscopic length, 
i.e. of the order of the lattice spacing in the lattice gas model.

Note that in the grandcanonical ensemble (temperature $T$ and  bulk field
$H_b$ chosen as control variables) liquid droplets in full thermal equilibrium
can exist neither in the bulk nor attached to the wall, even if a nonzero
bulk field favoring the liquid phase is switched on.  However, metastable droplets
can exist only for a finite "lifetime", and we shall address this case using concepts
of the theory of heterogeneous nucleation.  Thus, it is useful to recall that stable
liquid droplets of macroscopic size do exist when we consider the canonical ensemble,
choosing the density of the vapor $\rho$ in between the density of coexisting vapor
($\rho_v$) and liquid ($\rho_{\it l}$) phases.  The area fraction of the liquid then 
is given by the lever rule, $X = (\rho - \rho_v)/(\rho_{\it l} - \rho_v)$, and 
the shape of the liquid domain does depend on the choice of boundary conditions.
In our case, for small $X$ we obtain a wall attached droplet having the shape of
a circle cut with baseline $b_{drop}$ given in terms of the droplet radius $R$ and contact 
angle $\theta_c$ by 
\begin{equation} \label{A1}
b_{drop} = 2 R sin (\theta_c)
\end{equation}
and the contact angle for an isotropic interface tension $f_{int}$ between the 
coexisting vapor and liquid phases is given by Young\textquoteright s equation, namely
\begin{equation} \label{A2}
f_{int} cos(\theta_c) = f_{wv} - f_{w{\it l}}     \quad ,  
\end{equation}
where $f_{wv}$, $f_{w{\it l}}$ are the surface excess free energy   
densities of the vapor phase ($f_{wv}$) and liquid phase ($f_{w{\it l}}$)
due to the wall. 
Actually, Equation (\ref{A2}) is valid for a fluid in continuous space,
but not for the lattice gas model, where the interface tension $f_{int}(\theta)$
depends on the angle $\theta$ between the interface normal and the $x$ axis
of the lattice. Then Equation (\ref{A2}) needs to be replaced by   
\begin{equation} \label{A3}
f_{int}(\theta_c) cos(\theta_c) - sin(\theta_c) \frac{df_{int}(\theta_c)}{d\theta_c}|_{\theta = \theta_c}  = f_{wv} - f_{w{\it l}}     \quad . 
\end{equation} 

While in the isotropic case straightforward geometric considerations yield the area ($A$) of the 
circle cut as 
\begin{equation} \label{A4}
A = R^{2} (\theta_c -\frac{1}{2} sin(2 \theta_c)) = \frac{1}{4} b_{drop}^{2} (\frac{\theta_c}{sin^{2}(\theta_c)}  -\frac{cos(\theta_c)}{sin(\theta_c)})     \quad ,  
\end{equation}
\noindent 
and the length of the vapor-liquid interface line is
\begin{equation} \label{A5}
{\it l}_{{\it l}v} = 2 R \theta_c = b_{drop} \frac{\theta_c}{sin(\theta_c)}   \quad ,  
\end{equation}
\noindent finding the droplet shape for the anisotropic case is less straightforward.

In the bulk this problem is solved in terms of the well known Wulff construction, which for
the $d = 2$ Ising model can be worked out explicitly, and the shape of the droplet is given by
the equation \cite{60h,61h}
\begin{equation} \label{A6}
cosh(\tilde x) + cosh(\tilde y) = cosh(2J/k_B T)/tanh(2J/k_B T)   \quad ,  
\end{equation}
\noindent where $\tilde x$ and $\tilde y$ are the $x, y$ coordinates of the curve describing
the droplet shape. Equation (\ref{A6}) interpolates smoothly between a square shape
(for $T \rightarrow 0$) and a circle (for $T \rightarrow T_{cb}$). When we inscribe a circle
that touches the actual shape at $\tilde x = 0$ and at $\tilde y = 0$, it has a radius $R_{in}$
given by
\begin{equation} \label{A7}
R_{in} = arcosh [cosh(2J/k_B T)/tanh(2J/k_B T) -1 ]   \quad .  
\end{equation}
\noindent As an example we hence plotted Equation (\ref{A6}), 
in figure \ref{FigA1}, together with the inscribed circle of radius $R_{in}$  
evaluated at $t = T/T_{cb} = 0.40$, since for this choice of the reduced temperature
most of our simulations were made.
We found that the deviations from the spherical
shape are already rather minor, and this justifies our neglect of these anisotropy
effects, at least as a first approximation. 
The solution of Equation (\ref{A6}), reduces to the equation
of a circle near $T_{cb}$, where $\tilde x \rightarrow 0$ and $\tilde y \rightarrow 0$ and hence
\begin{equation} \label{A8}
\tilde x^2 + \tilde y^2 = 2 cosh(2J/k_B T)/tanh(2J/k_B T)  - 4 \quad ,  
\end{equation}
\noindent recalling that $cosh(2J/k_B T_{cb}) = 2^\frac{1}{2}$, $sinh(2J/k_B T_{cb}) = 1$, and hence 
$R_{in} \rightarrow 0$ as well. The solution for the wall-attached droplet then is given by the
Winterbotton construction \cite{62h}, i.e. we have to cut the droplet shown in figure \ref{FigA1}
by an horizontal straight line such that the angle of the tangent is $\theta_c$ as given by
Equation (\ref{A3}). The linear dimensions $\tilde y_{cut}$ and $R_{in}$ then follow from the 
condition that the area above the cut yields the desired area fraction $X$.

However, for the sake of simplicity we shall ignore these anisotropy effects in the following,
working with droplets of circular shape only. But even then there is one fundamental problem: there
is no physical reason for the $x-$coordinate of the center of mass of the droplet to coincide
with the origin of the coordinate system. In fact, this center of mass coordinate can be anywhere
on the $x-$axis when the boundary field is homogeneous, independent of $x$. Even in the
inhomogeneous case the droplets are only on average centered in the middle of the inhomogeneity,
as e.g. can be qualitatively observed in the snapshots of Figure \ref{fig2}. This
fact creates a translational entropy contribution $k_B T ln(M)$ for the droplet, where
$1 \le x \le M$ in our finite lattice of length $M$ in the homogeneous case.
Similar translational entropy contributions 
are known to hamper the numerical study of interfacial free energies \cite{64h,65h}. Thus, in 
a straightforward simulation study of the present problem the droplet would diffuse along the 
$x-$axis and its density profile $\rho(x,y)$ would be completely smeared out, until only the 
average translationally invariant density profile $\rho_{av}(y)$ is left, containing little
information on the droplet. Thus a "demon" would be needed to constrain the sampling of 
configurations such, that in each microstate of the system that is sampled the droplet center of mass
has its $x-$coordinate in the origin. Practical implementation of such a constraint 
is not completely trivial, since the size and the shape of the droplet due to their nanoscale 
dimensions is strongly fluctuating (c.f. figure \ref{fig2}).

We now consider the main subject of interest of the present paper, namely a boundary condition 
of the type shown in Figure \ref{fig1}, where the surface field $H_{w3}$ acts only over a distance 
$b$ along the $x-$axis, while in the remaining boundary a field $H_{w2} = -|H_{w3}|$ acts,
and hence the contact angle $\theta^{'}_{c} = \pi - \theta_{c}$ applies.

In $d = 3$ dimensions in the canonical ensemble, this situation has already been 
considered by Lipowsky et al. \cite{66h,67h,68h}.
They pointed out that three regimes
need to be distinguished, namely: (I) $b_{drop} < b$; (II) $b_{drop} = b$; and (III) $b_{drop} > b$, 
see Figure \ref{FigA2} adapted to our $d = 2$ dimensional case. Here the area $A$ of the
wall attached droplet is the control parameter that is varied: For sufficiently small area taken 
by the liquid baseline $b_{drop}$ that will result from $A$ and $\theta_c$ via Equation (\ref{A4}) will
be in the regime I, and the $x-$coordinate of the center of mass of the droplet can be
anywhere in the interval from $x = -(b-b_{drop})/2$  to   $x = +(b-b_{drop})/2$. Unlike Lipowsky   
et al. \cite{66h,67h,68h} we do not assume that the droplet is exactly centered at $x = 0$,
the center of the inhomogeneity of the wall, which is our origin. This center certainly is
the most probable position, but there will be a broad probability distribution for this
center of mass coordinate, and when we consider the average density profile  $\rho(x,y)$ 
obtained by convoluting the density profile of the droplet with baseline $b_{drop}$ and contact angle
$\theta_c$ with this probability distribution, a density distribution $\rho_{ave}(x,y)$ must result
that is considerably flattened in comparison with $\rho(x,y)$.  From $\rho_{ave}(x,y)$ one would obtain 
an effective contact angle $\theta_c^{eff}$ that clearly will be much smaller than the correct one,
if $b_{drop} \ll b$. This entropic effect was disregarded  by Lipowsky et al. \cite{66h,67h,68h}, but clearly 
must be present in our simulations and thus hampers their interpretation.
It is tempting to associate the small values of $\theta_{eff}$ in figures \ref{fig6}  and \ref{fig7} observed
for ${\it l}_{eff} < b$ with this flattened profiles due to the fluctuations in the
center of mass position of small wall-attached droplets.

The most interesting situations of course, are found when $b_{drop}$ as given by Equations (\ref{A1}),
(\ref{A4}) has reached the value  $b_{drop} = b$: then the prediction is that further increase of $A$ 
does not cause a further growth of $b_{drop}$. Rather, what happens is a growth of the contact angle 
$\theta$ of the droplet from the value $\theta_c$ given by the Young's equation to a larger value,
satisfying an equation analogous to Equation (\ref{A4}), namely
\begin{equation} \label{A9}
A = \frac{1}{4} b^{2} (\frac{\theta}{sin^{2}(\theta)}  -\frac{cos(\theta)}{sin(\theta)} ) \quad ,
\theta_c < \theta < \theta_c^{'}  \quad . 
\end{equation}
\noindent Thus in a sense the interface between liquid and vapor is pinned at the points 
$x = \pm b/2$ when $A$ has increased up to the value where Equation (\ref{A9}) yields
$\theta =  \theta_c^{'}$ ($= \pi - \theta_c$, in our case),
depinning of the interface from the inhomogeneities of the boundary occurs, and 
$\theta$ stays at $\theta_c^{'}$, while $b_{drop} > b$. Again Lipowsky et al. \cite{66h,67h,68h}
have assumed that the $x-$coordinate of the center of mass of the droplet is at $x = 0$, but
we maintain that again fluctuations will occur. However, the 
region $b_{drop}$ in between the two contact points of the interface will always encompass the 
region of the inhomogeneity, from $x = -b/2$ to $x = +b/2$, and the average position of the 
center of mass of the droplet will hence have $x-$coordinate $x = 0$.

When we now turn to the description in the grandcanonical ensemble, we note that a correspondence
to the droplet configurations discussed for the canonical ensemble can exist only when the droplet
configurations in the grandcanonical ensemble are still metastable.

For the problem without boundary inhomogeneity we have the standard problem of heterogeneous
nucleation at the wall. The free energy cost of the forming droplet is written as the 
excess free energy relative to the wall without droplet, namely
\begin{equation} \label{A10}
\Delta F_{drop} = -2 m_{coex} H_b A + \Delta F_{int}
\end{equation}
\noindent where $m_{coex}$ is the spontaneous magnetization of the Ising model,
and $A$ is given by Equation (\ref{A4}) and $\Delta F_{int}$ becomes,
\begin{equation} \label{A11}
\Delta F_{int} = 2 R f_{int} \theta_c + (f_{wl} - f_{wv}) 2 R sin(\theta_c) = 
2 f_{int} R[\theta_c - \frac{1}{2} sin(2 \theta_c)]   \quad ,
\end{equation} 
\noindent where Equation (\ref{A2}) was used. Equations (\ref{A10}) and (\ref{A11}) yield
\begin{equation} \label{A12}
\Delta F_{drop} = [\theta_c - \frac{1}{2} sin(2 \theta_c)][- 2 m_{coex} H_b R^{2} + 2 f_{int} R].
\end{equation} 

Minimizing $\Delta F_{drop}$ with respect to $R$ yields
\begin{equation} \label{A13}
R^{*} = f_{int} / (2 m_{coex} H_b)  \quad , \Delta F^{*}_{het} = \Delta F^{*}_{homo} f_{VT} (\theta_c)  \quad ,
\end{equation} 
\noindent where $\Delta F^{*}_{homo}$ is the standard result for the free energy barrier 
against homogeneous nucleation in $d = 2$ dimensions  
\begin{equation} \label{A14}
\Delta F^{*}_{homo} = \frac{\pi}{2} f_{int}^{2} /(m_{coex} H_b)   \quad ,
\end{equation} 
\noindent and $ f_{VT} (\theta_c)$ is the analog of the well-known Volmer-Turnbull
function in $d = 2$ dimensions, given by
\begin{equation} \label{A15}
f_{VT} (\theta_c) = \frac{1}{\pi}  [\theta_c - \frac{1}{2} sin(2 \theta_c)] \quad .
\end{equation}

Note that $f_{VT} (\theta_c) \approx \frac{2}{3 \pi} \theta_c^{3}$ for $\theta_c \rightarrow 0$,
when complete wetting begins.  It turns out, of course, that use of $R^{*} = f_{int} / (2 m_{coex} H_b) $
in Equation (\ref{A1}) yields $b_{drop} \ll b$ only for rather large fields. All the data where the metastable
droplets are encountered do not fall in this regime, as expected.

\section{Pinned droplets: Comparing the two- and the three-dimensional cases.}

While the numerical simulation work exclusively has addressed the case of a
two-dimensional system with a one-dimensional boundary where the positive surface
field (favoring the liquid phase of the lattice gas) acts on a length $b$,
it is also instructive to consider the three-dimensional case, where the positive
surface field acts on a circular heterogeneity with radius $r$.
For the sake of clarity, the geometry of the pinned droplet is sketched in Figure \ref{figB1}.

The radius of curvature of the sphere-cap shaped droplet is $R$. Then 

\begin{equation} \label{PD1}
r = R sin (\theta), \quad    h = R(1 - cos(\theta))  \quad ,
\end{equation}
  
\noindent where it is convenient to express all quantities in terms of the
height $h$ of the droplet above the substrate. The angle $\theta$ that the droplet
makes with the substrate can be in the range 

\begin{equation} \label{DP2}
  \theta_c \leq \theta \leq \pi - \theta_c   \quad ,
\end{equation}
  
\noindent where $\theta_c$ is the contact angle given by Young's equation. Notice that
only for angles in the quoted range droplets with basal radius $r$ exist; however,
only for $\theta \leq \pi/2$ such droplets are metastable, while for
$\pi /2 \leq \theta \leq \pi - \theta_c$ they are  unstable.

Now the volume of the sphere cap is

\begin{equation} \label{DP3}
  V = \frac{\pi h}{6} (3 r^2 + h^2) \quad ,
\end{equation}   
\noindent also the basis surface is $\pi r^2$, while the upper surface is 

\begin{equation} \label{DP4}
A_u = \pi (r^2 + h^2) \quad .
\end{equation} 

So, the free energy difference of the droplet of height $h$
relative to a disk-shaped droplet of radius $r$ and height $h = 0$ is
(the choice of this reference state is arbitrary, of course)

\begin{equation} \label{DP5}
\Delta F =  f_{int} A_u  - 2 m_{coex} H_b V  \quad ,
\end{equation}
  
\noindent where $H_b$ is the bulk field. Then, by using Equations (\ref{DP3})
and (\ref{DP4}) one obtains 

\begin{equation} \label{DP6}
\Delta F =  f_{int} \pi (r^2 + h^2)  -  m_{coex} H_b \frac{\pi}{3} h (3 r^2 + h^2) \quad .
\end{equation}

It is convenient to find the extrema of $\Delta F$ simply as a function of $h$; then

\begin{equation} \label{DP7}
(\partial (\Delta F) / \partial h)_{H_b} = 0  \quad ,
\end{equation} 

\noindent yields

\begin{equation} \label{DP8}
  h^2 - 2h \frac{f_{int}}{m_{coex} H_b}  + r^2 = 0    \quad  ,
\end{equation}  

\noindent such that in terms of $\tilde H_b = H_b r m_{coex} / f_{int}$
one finds for $\tilde H_b < 1$ two solutions, namely

\begin{equation} \label{DP9}
\frac{h}{r} = \tilde H_b^{-1}  \pm \sqrt{(\tilde H_b^{-2} - 1)} \quad .
\end{equation}

\noindent The minus sign yields the free energy minimum, corresponding to
the pinned droplet, while the plus sign corresponds to a surface
free energy maximum, and the corresponding angle $\theta$ can be read off from
\begin{equation} \label{DP10}
tan(\frac{\theta}{2}) = \frac{1 - cos(\theta)}{sin(\theta)} = \frac{h}{r}   \quad .
\end{equation}
  
The limiting case $\tilde H_b = 1$ means $h/r = 1$, $\theta = \pi/2$, i.e.
a semispherical droplet. The free energy function can be written as
\begin{equation} \label{DP11}
\Delta F /  f_{int} \pi  r^2 = \frac{2}{3} (\tilde H_b^{-2} \pm \tilde H_b^{-1} \sqrt{(\tilde H_b^{-2} - 1)} \mp \tilde H_b \sqrt{(\tilde H_b^{-2} - 1)} )  \quad ,
\end{equation}
\noindent and hence the barrier for the depinning transition of the pinned droplet becomes
\begin{equation} \label{DP12}
  \Delta F_2 /  f_{int} \pi  r^2 = \frac{4}{3}  \tilde H_b (\tilde H_b^{-2} -1)^{3/2} \quad .
\end{equation}   
\noindent From this calculation it is obvious that the mathematics in $d = 3$
is even simpler than in $d = 2$, since the use of $h$ instead of the angle
$\theta$ makes the description of  $\Delta F$  (c.f. Equation (\ref{DP5}))
very simple. In $d = 2$, equation (\ref{DP10}) also holds, but equation (\ref{A16})
shows that both $\theta$ and $sin(2\theta)$ enter in the free energy expression,
so no simple formula for $\Delta F(h)$ in $d = 2$ can be written down. When one works
out $h/r$ and $\Delta F_2$ in both $d = 2$ and $d = 3$, one notes a very similar behavior:
near the point  $\tilde H_b = 1$ the barrier vanishes like
$(1 - \tilde H_b^{-1})^{3/2}$, i.e. with a vanishing slope, and $h/r$ reaches the semicircle or semisphere
configuration with a square-root cusp.
Figure \ref{figB2} shows a comparison of the free energy barrier of the
depinning transition of pinned droplets for the two- and the three-dimensional cases.

\underline{Acknowledgments}: E.V.A. is grateful to the Alexander von Humboldt foundation and 
to the Deutsche Forschungsgemeinschaft (DFG, SFB TRR 146) for partial support of his research 
stays at the Institut f\"ur Physik of the Johannes Gutenberg Universit\"at Mainz.
Also, M.L.T. and E.V.A are grateful to the CONICET and UNLP (Argentina) for financial
support.


\end{document}